\documentclass[
reprint,
amsmath,amssymb,
aps,
prb,
]{revtex4-1}

\usepackage{graphicx}
\usepackage{dcolumn}
\usepackage{bm}
\usepackage{amsmath,amsfonts}
\usepackage{floatrow}
\usepackage{xcolor}
\usepackage{placeins}

\usepackage{comment}
\begin{document}

\title{Functional universality in slow-growing microbial communities arises from thermodynamic constraints}
\author{Ashish B. George$^{1,2}$, Tong Wang$^{1,3,4}$, and Sergei Maslov$^{1,3,5\ast}$ }

\affiliation{
\vskip 10pt
$^1$Carl R. Woese Institute for Genomic Biology, University of Illinois at Urbana-Champaign, Urbana, IL 61801, USA.\\
$^2$Department of Plant Biology, University of Illinois at Urbana-Champaign, Urbana, IL 61801, USA.\\
$^3$Department of Physics, University of Illinois at Urbana-Champaign, Urbana, IL 61801, USA.\\
$^4$Brigham and Women’s Hospital and Harvard Medical School, Boston, MA 02115, USA.\\
$^5$ Department of Bioengineering, University of Illinois at Urbana-Champaign, Urbana, IL 61801, USA.}
\author{\footnotesize{$^\ast$~\url{maslov@illinois.edu}}}

\date{\today}

\date{\today}

\begin{abstract}
\noindent The dynamics of microbial communities is incredibly complex, determined by competition for metabolic substrates and cross-feeding of byproducts. Species in the community grow by harvesting energy from chemical reactions that transform substrates to products. In many anoxic environments, these reactions are close to thermodynamic equilibrium and growth is slow. To understand the community structure in these energy-limited environments, we developed a microbial community consumer-resource model incorporating energetic and thermodynamic constraints on an interconnected metabolic network. The central ingredient of the model is product inhibition, meaning that microbial growth may be limited not only by depletion of metabolic substrates but also by accumulation of products. We demonstrate that these additional constraints on microbial growth cause a convergence in the structure and function of the community metabolic network---independent of species composition and biochemical details---providing a possible explanation for convergence of community function despite taxonomic variation observed in many natural and industrial environments. Furthermore, we discovered that the structure of community metabolic network is governed by the thermodynamic principle of maximum heat dissipation. Overall, the work demonstrates how universal thermodynamic principles may constrain community metabolism and explain observed functional convergence in microbial communities.
\end{abstract}

\maketitle

\section*{Introduction}

Over half of earth's prokaryotic biomass resides in ocean sediments and deep soil~\cite{falkowski_microbial_2008,whitman_prokaryotes_1998}. In many of these natural environments and anaerobic bioreactors, communities grow under energy-limited conditions due to the absence of strong electron acceptors, like oxygen or nitrate, and strong electron donors, like glucose~\cite{jessen_hypoxia_2017, jorgensen_feast_2007, leng_review_2018, hoh_experimental_1997, nobu_catabolism_2020, bradley_widespread_2020}. Hence species in these communities are forced to harvest energy required for biomass growth from low free energy reactions~\cite{conrad_thermodynamics_1986, bradley_widespread_2020, hickey_thermodynamics_1991, jackson_anaerobic_2002,larowe_energetics_2019,larowe_thermodynamic_2012}. Growth in these environments can be very slow---division times are measured in weeks and years---instead of minutes and hours used for growth in the lab \cite{larowe_thermodynamic_2012,lomstein_endospore_2012,jorgensen_feast_2007,park_impact_2010,sottile_comparative_1977}. These long timescales and difficulty in culturing these species in a lab environment make experimental investigation of these communities challenging. Models and theory of energy-limited microbial communities can help uncover their organizational principles and shared emergent properties and drive understanding.

Models and theories inspired by MacArthur's consumer-resource model have helped understand many of the principles and properties governing microbial communities structured by nutrient-limitation~\cite{macarthur_species_1970, posfai_metabolic_2017,tikhonov_collective_2017, goldford_emergent_2018,marsland_minimum_2020, dubinkina_multistability_2019, wang_complementary_2021}. However, models of communities in low-energy environments, structured by energy-limitation, need to incorporate important additional details. First, microbial growth in low-energy environments is determined by energy assimilation rather than nutrient uptake. Second, the low free energy reactions utilized to drive growth are subject to thermodynamic constraints. And third, reactions need to be explicitly modeled to track their thermodynamic feasibility, rather than just species preference for resources. Here we propose and study a minimal model of microbial communities that accounts for these crucial details to understand the emergent properties of slow-growing communities in energy-limited environments. 

Recently, a few studies have attempted to incorporate thermodynamics into community models~\cite{groskopf_microbial_2016, cook_thermodynamic_2021}. These studies focused on community diversity, showing that thermodynamics enables communities to overcome the competitive exclusion principle and stabilize high diversity on a single substrate. In contrast, here we focus on the emergent properties of the communities in low-energy environments, such as the metabolic network structure and function of the community.

 Simulations of our model generated communities that shared many features of natural communities. Strikingly, communities assembled from separate species pools in replicate environments realized the same metabolic network and performed similar metabolic functions at the community-level despite stark differences at the species level. This resembles the observed convergence in community function despite taxonomic divergence, seen in anaerobic bioreactors, oceans, and other environments~\cite{fernandez_how_1999, louca_decoupling_2016,louca_function_2018, peces_deterministic_2018}. 
 
 Analyzing the model to understand how functional convergence arises, we discovered that thermodynamic principles govern community-level metabolic network structure and function in slow-growing, energy-limited communities. We derived the thermodynamic principle of maximum dissipation, which, we show, determines the community metabolic network selected by ecological competition in energy-limited communities. This thermodynamic principle, along with physical constraints, shape the metabolic environment created by the community and the metabolic functions it performs. Further, the derivation of the maximum dissipation principle provides a concrete example in which communities are structured by thermodynamic optimization, an idea that has been conjectured in ecology but never proven~\cite{dewar_theoretical_2014}.

Overall, our results highlight the crucial role of thermodynamic constraints in defining community metabolism and function of energy-limited microbial communities. The results provide an explanation for the observed convergence in microbial community function, serve as a concrete example of thermodynamic optimization in community ecology, and make predictions for the community metabolic network and resource environment.

\section*{Results}
\subsection*{Model of microbial communities in energy-poor environments}

\begin{figure*}
\includegraphics[width=\textwidth]{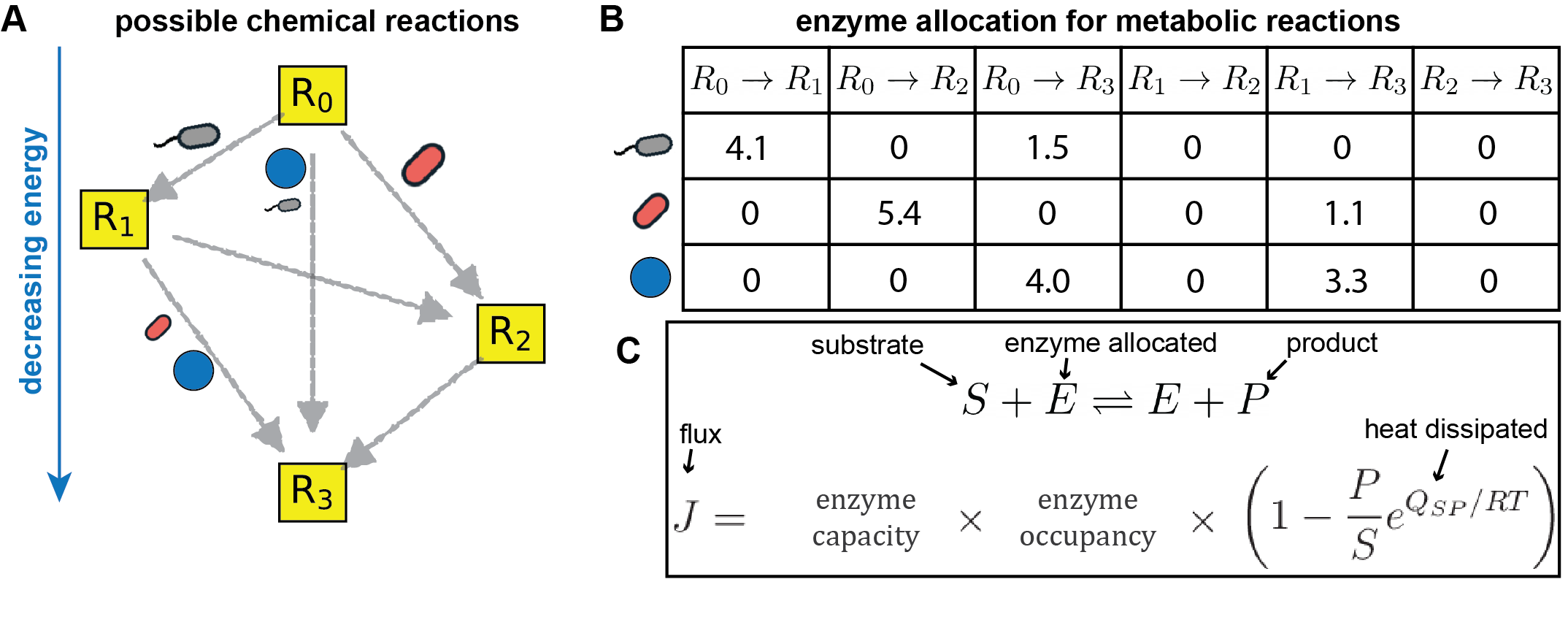}
\caption{\textbf{Model of communities in energy-limited environments.} 
\textbf{A)} Species harvest energy from reactions (grey arrows) that consume higher energy resource (substrate) to produce a lower energy resource (product). Reactions connecting any pair of the four resources $R_0,R_1,R_2,R_3$ can be utilized by a species, making the space of all possible reactions a fully connected network. 
\textbf{B)} A species can utilize a reaction by allocating a part of its overall enzyme budget to the corresponding, reaction-specific enzyme. Species differ in both the manner of enzyme allocation as well as their overall enzyme budgets. 
\textbf{C)} The reactions in low-energy environments are thermodynamically reversible. The flux $J$ in a reversible, enzyme-catalyzed reaction between substrate $S$ and product $P$ is determined by three factors: 1) Enzyme capacity, which is the amount of enzyme available. 2) Enzyme occupancy, which measure the fraction of free enzyme available to catalyze the reaction, determined by the concentration of substrates and products. 3) Thermodynamic inhibition due to accumulation of products. Since low-energy reactions are reversible, the net flux in the forward direction will decrease as products accumulate. The strength of this inhibition is controlled by the heat dissipated in the reaction, $Q$. }
\label{fig:model_cartoon}
\end{figure*}

To understand slow-growing microbial communities in energy-limited environments, we develop a model that incorporates crucial features of growth in these environments: energetic-limitation of growth, thermodynamic constraints on reactions, and explicit modeling of reaction fluxes.

In the model, species grow utilizing energy harvested from chemical reactions that convert a higher energy resource (substrate) to a lower energy resource (product) (Fig.~\ref{fig:model_cartoon}A). A reaction can be understood as a coarse-grained description of catabolism~\cite{jin_thermodynamics_2007,wachtel_thermodynamically_2018}. The resources are labeled in the order of decreasing standard state energy as $R_0,R_1,R_2,...$. A species can utilize reactions connecting any pair of resources; the allowed reactions make a directed, fully-connected network. The product of one reaction can act as the substrate of another, facilitating cross-feeding interactions between microbes. 

Species utilize a reaction by producing the corresponding, reaction-specific enzyme. A species allocates a portion of its total enzyme budget to any number of the possible reactions. We allow total enzyme budgets to vary, as in real species~\cite{milo_cell_2016}; this also avoids potential degenerate behavior seen in community models with fixed enzyme budgets~\cite{posfai_metabolic_2017}. Thus a species is characterised by its enzyme allocation strategy and total enzyme budget (Fig.~\ref{fig:model_cartoon}B). Through the explicit connection of species to reactions, substrate consumption can create different products depending on the enzyme used, like in real microbes~\cite{krieger_metacyc_2004}. This contrasts with microbial consumer-resource models, where substrate consumption by a species leads to a fixed mixture of products, i.e., all species use the same reactions and differ only in their consumption preferences~\cite{marsland_minimum_2020,marsland_available_2019,fant_eco-evolutionary_2021}.

From a reaction converting substrate $S$ to product $P$ with standard-state energies $\mathcal{E}_S^0$ and $\mathcal{E}_P^0$, a species assimilates $\mathcal{E}^{\mathrm{ATP}}$ energy. This energy, stored in ATP or other forms, is constrained by reaction stoichiometry, and so is assumed to be species-independent.
 
The harvested energy drives the proportional growth of new biomass. The remaining energy is dissipated as heat $Q_{SP}$. $Q_{SP}$ is is defined as 
\begin{equation}
Q_{SP}=   \mathcal{E}_S^0 -\mathcal{E}^{\mathrm{ATP}} -\mathcal{E}_P^0,
\end{equation}
which can also be understood as the negative free energy difference under standard conditions, $\Delta G^{0}$.

The flux through the reaction is determined by reversible Michaelis-Menten kinetics~\cite{cornish-bowden_fundamentals_2012} : 
\begin{equation}
J_{S\to P}= k_{\mathrm{cat}} E \frac{ \frac{S}{K_S}   }{1 +  \frac{S}{K_S} + \frac{P}{K_P}  } \left(1- \frac{P}{S} \; \mathrm{exp}\left[-\frac{Q_{SP}}{RT} \right]
\right),
\label{eq:JSP}
\end{equation}
where $E$ is the enzyme concentration, $k_{\mathrm{cat}}, K_S, K_P$ are enzyme-specific parameters describing the chemical kinetics, $T$ is the temperature, and $R$ is the gas constant. In thermodynamic parlance, $\frac{P}{S} \; \mathrm{exp}\left[-\frac{Q_{SP}}{RT}\right]$ represents the free energy difference after accounting for resource concentration gradient, $\Delta G^{\prime}=\Delta G^{0} +RT \log \frac{P}{S}$. Note that our results will apply to a general stoichiometric ratio of substrate and product beyond $1:1$ as well~(see SI); we omit discussion of this scenario in the main text for clarity. 

Eq.~\ref{eq:JSP} can be understood as a product of three terms~\cite{noor_note_2013}. The first term specifies the maximal reaction rate, which is proportional to amount of enzyme allocated to the reaction, $E$. The second term describes the saturation level of the enzyme; the flux increases linearly at low substrate concentrations and saturates at high substrate concentrations. The last (and most crucial for our purposes) term describes thermodynamic inhibition, i.e., how the net rate of a reversible reaction decreases as products accumulate and the ratio $P/S$ increases. The strength of the thermodynamic inhibition is controlled by the heat dissipated in the reaction. This thermodynamic inhibition of microbial growth has been quantitatively validated in the lab and in situ conditions~\cite{jin_new_2003,hoehler_thermodynamic_1998,hoh_experimental_1997,westermann_effect_1994}. If product accumulation is severe enough to reverse the reaction direction, we assume that species down-regulate enzyme production to prevent energy loss from catalyzing the reverse reaction. 

The dynamics of species and resources are driven by these reaction fluxes. Species grow in proportion to the energy they assimilate from the reactions catalyzed by their enzymes. The concentration of a resource decreases due to consumption in a reaction and increases due to production in a reaction. Additionally, the entire system is diluted at rate $\delta$, which represents the dilution rate of the bioreactor or the dilution by sedimentation in oceans~\cite{arndt_quantifying_2013}. This dilution of species biomass can also be extended to account for the maintenance energy requirements of a species (see SI). The precise equations describing species and resource dynamics are detailed in Methods. We also obtained similar results in a model without explicit dilution of resources and only species maintenance costs(see SI); we omit discussion of this model in the main text for clarity.

Communities in natural and industrial environments are the result of years of ecological competition and succession. To model such communities that emerge from ecological competition, we studied community assembly from a large and diverse species pool in an environment consisting of $6$ resources, labelled in the order of decreasing standard-state energies as $R_0, R_1..., R_5$. The environment was supplied with the most energy-rich resource $R_0$ at high concentrations compared to the typical $K_S$. The other lower energy resources were generated as products in some of the 15 reactions that could be utilized by species for growth. The energy assimilated from a reaction was chosen to be a random fraction (between $15\%$ and $85\%$) of the energy gap between product and substrate of that reaction. Enzymatic parameters describing reaction kinetics, were chosen from lognormal distributions, motivated by empirical observations~\cite{bar-even_moderately_2011}. Since we are interested in slow-growing communities, we chose a small dilution rate so as to not drive any slow-growing species to extinction. After introducing species from the pool into the environment, many species went extinct from competition and simulations settled down to a steady-state community comprised of the survivors.

\subsection*{Emergent metabolic structure and function in slow-growing, energy-limited communities }

To study the features shared across slow-growing communities in energy-limited environments, we repeated the community assembly experiment from five separate pools of $600$ species each. Each species allocated its enzyme budget, chosen from a lognormal distribution, to a random subset of the 15 possible reactions. Species were not shared across pools. We also varied enzymatic parameters (chemical kinetic parameters) to vary between pools. The environment and its energetics was shared across the experiments, i.e., the resource supply, resource energies, and energy uptake in each reaction (which is constrained by stoichiometry) was the same across experiments. Simulations of all pools converged to a steady-state maintained by a community of the surviving species.

\begin{figure*}
\includegraphics[width=.75\textwidth]{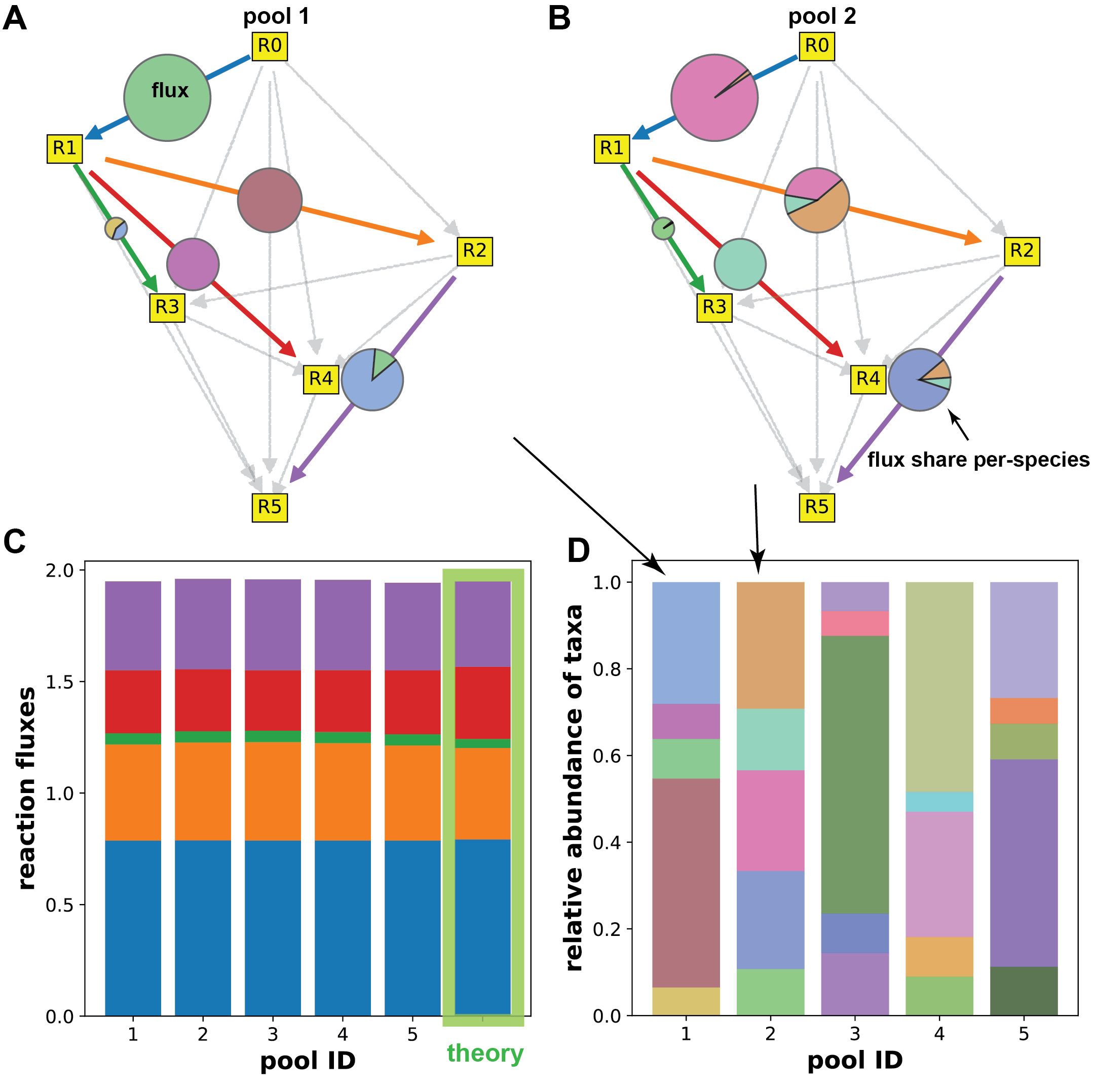}
\caption{\textbf{Functional convergence despite taxonomic variation.} 
We simulated the model to obtain the final community that arose as the outcome of competition between a large number of species. We repeated this community assembly experiment from five separate pools of $600$ species each.
\textbf{A,B)} The metabolic networks in the final communities obtained from two separate species pools are similar. From all possible reactions (grey arrows), the same subset of reactions are active (i.e., carry a positive flux) in the final communities (colored arrows). On each reaction arrow, the size of the circles show the flux carried in the reaction and the pie-chart on shows how this flux is shared between species. The manner in which species contribute to the flux differs between pools.
\textbf{C)} The flux through the community metabolic network is similar across communities from different species pools. The theoretical prediction (Eq.~\eqref{eq:Ri_ss}) matches simulations. The reactions in the bar plot are colored according to the arrow colors in panels A,B (pools 1 and 2). 
\textbf{D)} The relative abundance of species in the final communities. The species in the bar plot colors are colored according to the pie-charts in panels A,B (pools 1 and 2). The relative abundance is not proportional to the flux catalyzed by the species because the energy-yield per unit flux differs between reactions. For example, the green species from pool 1 (panel A) and the pink species from pool2 (panel B) catalyze a large flux but have a low relative abundance due to the low energy yield of the reaction (panel C). }
\label{fig:functional_convergence}
\end{figure*}

We analyzed the final steady-state communities to find emergent properties of energy-limited communities that were shared across pools that differed in species and enzyme content. We identified three properties of the communities that resemble observations in natural communities:

\textbf{Concurrent community metabolic networks:} Only a subset of the 15 possible reactions were active (i.e., carried nonzero flux) in the final communities (colored arrows in Fig.~\ref{fig:functional_convergence}A). This subset of reactions was shared across communities, despite starting from different species pools (Fig.~\ref{fig:functional_convergence}A,B). Hence, the reactions utilized by the surviving species were determined by the environment. Furthermore, the metabolic network structure was constrained. The in-degree of each node was $1$, i.e., each resource was produced by only a single reaction. 

\textbf{Functional convergence:} We compared the metabolic functioning of the communities obtained from the different species pools. The function performed by the community was defined as the flux through the active reactions in the final community. These fluxes were strikingly similar across pools~(radius of circles in Fig.~\ref{fig:functional_convergence}A,B and Fig.~\ref{fig:functional_convergence}C). Thus communities converged in the metabolic function they performed despite having been assembled from different species pools.

\textbf{Taxonomic divergence:} Bolstered by the convergence in metabolic network structure and function across pools, we measured the species abundances and functional roles in the communities. The functional role of a species was quantified as the reaction flux catalyzed by it. We found that species contributed to the reaction fluxes idiosyncratically across pools~(pie charts in Fig.~\ref{fig:functional_convergence}A,B). Some reactions were performed by multiple species while others were performed by only one, and this breakup of reaction fluxes between the species varied between pools. The relative abundance of the species is determined by product of the metabolic flux catalyzed by a species and the energy assimilated per unit flux, which differed between reactions. The relative abundance is shown in Fig.~\ref{fig:functional_convergence}D. Thus the relative abundance and functional roles of the species did not converge across the different pools.

Taken together, these observations mirror the `functional convergence despite taxonomic divergence' observed in many natural communities~\cite{fernandez_how_1999, louca_decoupling_2016,louca_function_2018}. We investigated the model analytically to understand why these slow-growing, energy-limited communities share these emergent properties. We uncovered fundamental thermodynamic principles governing functional convergence in these energy-limited communities, which we explain in the next section.

\subsection*{Principle of maximum heat dissipation explains selected community metabolic network}

We first sought to understand why the same metabolic reactions are realized in the final community across species pools. We were able to make analytical progress by focusing our attention on slow-growing communities growing on low-energy reactions. At steady-state, the biomass growth in communities replaces loss by dilution, $\delta$. In slow-growing communities, the growth at steady-state is much smaller than the maximum growth rate of the species $g_{\mathrm{max}}$. Motivated by conditions in ocean sediments and bioreactors, the substrate concentration in the resource supply in simulations was larger than the typical $K_S$~\cite{mentges_long-term_2019, arndt_quantifying_2013,jessen_hypoxia_2017,fernandez_how_1999}. The high substrate levels in the supply mean that growth was primarily slowed by the accumulation of products.

\begin{figure}
\includegraphics[width=\textwidth]{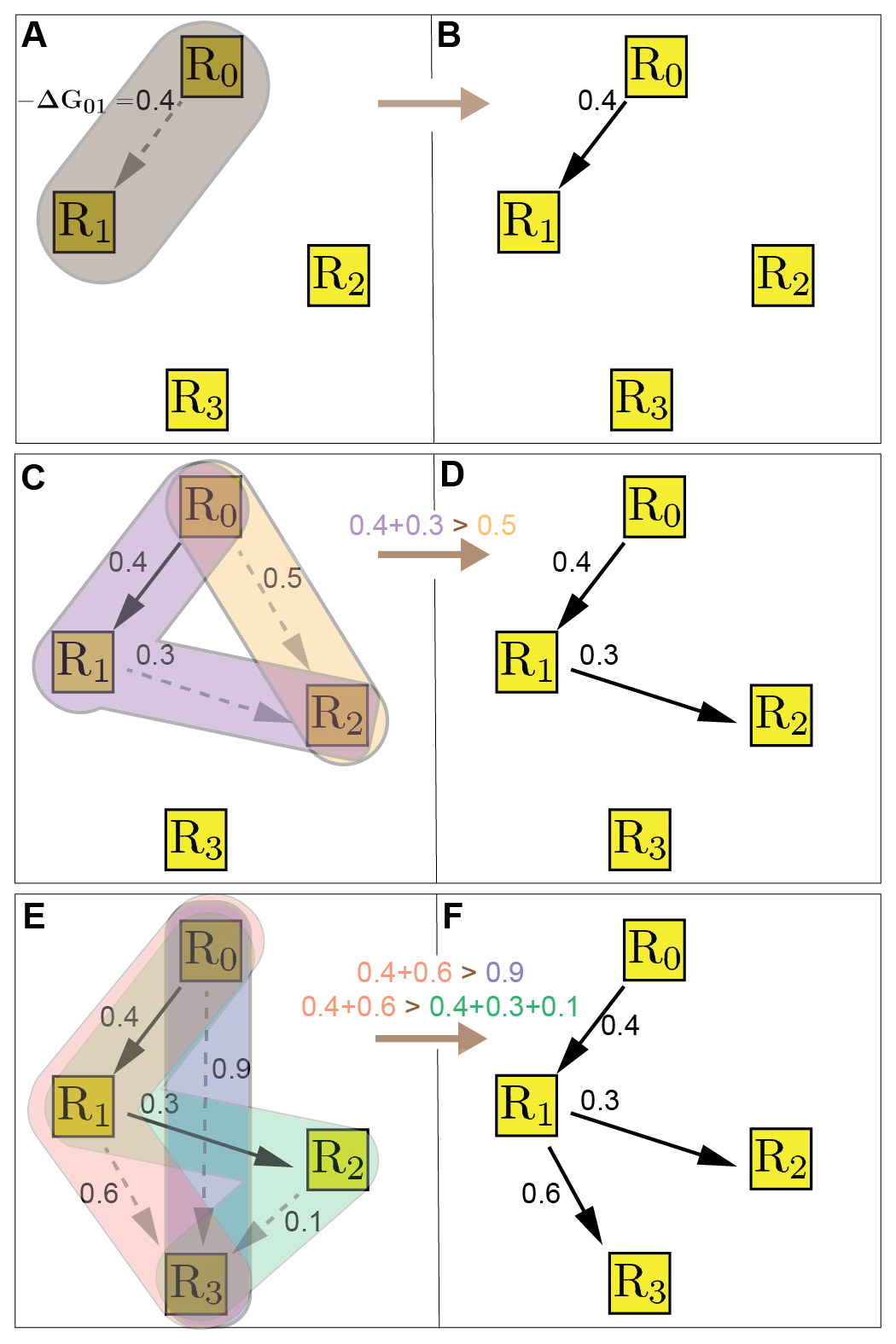}
\caption{\textbf{The maximum dissipation principle determines the community metabolic network.} In a system with four resources, we assemble the community metabolic network by considering the production of each resource sequentially to understand the governing thermodynamic principles.
\textbf{A)} The sole path from the supplied resource, $R_0$, to the next-highest energy resource $R_1$. The path dissipates $Q_{01}=0.4$ units of energy as heat. 
\textbf{B)} The steady-state community with reaction $R_0 \to R_1$ maintains resource concentrations such that $\frac{R^*_1}{R^*_0} \approx \mathrm{exp}\left(0.4/RT\right) $.
\textbf{C)} The two possible paths from $R_0$ to the next resource, $R_2$, that could be added to the community. The paths $R_0 \to R_2$ and $R_0 \to R_1 \to R_2$ dissipate 0.5 and 0.7 units of energy as heat. 
\textbf{D)} The path that dissipates more heat, $R_0 \to R_1 \to R_2$, is selected in the final community because species utilizing this path will drive concentration of $R_2$ higher (such that $\frac{R^*_2}{R^*_0} \approx  \mathrm{exp}\left(0.7/RT\right) $) than species utilizing $R_0 \to R_2$
\textbf{E)} The three possible paths from $R_0$ to $R_3$, that could be added to the community. We do not consider the path $R_0 \to R_2 \to R_3$ because we have already found that the maximally dissipative path to $R_2$ is $R_0 \to R_1 \to R_2$ and not $R_0 \to R_2$.
\textbf{F)}The path that dissipates most heat, $R_0 \to R_1 \to R_3$, is selected in the final community because it can drive the concentration gradient $R_3/R_0$ highest (such that $\frac{R^*_3}{R^*_0} \approx  \mathrm{exp}\left(1.0/RT\right)$).
Thus, the active reactions in the final community network will be determined by the paths to each resource that dissipates the most heat.}
\label{fig:max_heat_dissip}
\end{figure}

Since growth is slowed down by product accumulation, we can relate the concentrations of substrates and products of an active reaction in the final steady-state community. Fig.~\ref{fig:max_heat_dissip}A shows an example of a reaction $R_0 \to R_1$ in a community with four resources. If the reaction is active in the final steady-state community, then we have:
\begin{equation}
    \frac{R^*_1}{R^*_0} = e^{Q_{01}/RT}  \;  + \mathcal{O}\left(\frac{\delta}{g_{\mathrm{max}}}\right),
    \label{eq:R1_R0star}
\end{equation}
 where $R^*_1,R^*_0$ are the steady state concentrations of the resources and $Q_{01}$ is the heat dissipated in the reaction $R_0 \to R_1$. This condition implies that reactions are slowed down to near thermodynamic equilibrium due to product accumulation. The first (and leading order) term depends only on the thermodynamics of the reaction. The second term, $\frac{\delta}{g_{\mathrm{max}}}$, which contains all the properties specific to the species and enzyme, is only a small correction.

We now consider the production of the second resource, $R_2$, which can be produced via two possible reaction paths~(Fig.~\ref{fig:max_heat_dissip}C). To analyze this scenario, we derive a crucial extension to Eq.\eqref{eq:R1_R0star} that applies to reaction paths, composed of multiple reactions, that are active in the final community~(see SI for derivation). If a reaction path proceeding from $R_i$ to $R_j$ is active in the final community, we have:
\begin{equation}
    \frac{R^*_j}{R^*_i} = e^{Q_{i \Rightarrow j}/RT}  \;  + \mathcal{O}\left(\frac{\delta}{g_{\mathrm{max}}}\right),
    \label{eq:RiRj_at_SS}
\end{equation}
where $Q_{i \Rightarrow j}$ refers to the sum of the heat dissipated in the reactions along the path from  $R_i$ to $R_j$. Again, note that the species- and enzyme- specific properties are only a small correction. This equation provides us different values for $\frac{R^*_2}{R^*_0}$ for each of the paths considered in Fig.~\ref{fig:max_heat_dissip}C. Since both values cannot hold simultaneously, only one of the two paths can be realized in the final community. Thus the in-degree of any node in the network is always $1$, as seen in Fig.~\ref{fig:functional_convergence}A,B.

To understand which of the two paths is selected in the final community, we can consider the competition of two species utilizing reactions $R_1 \to R_2$ and $R_0 \to R_2$ added to the community with reaction $R_0 \to R_1$ shown in Fig.~\ref{fig:max_heat_dissip}B. The species utilizing the reaction $R_1 \to R_2$ (and hence path $R_0 \to R_1 \to R_2$) will drive product accumulation until $\frac{R^*_2}{R^*_0} \approx e^{0.7/RT}$, while the species utilizing the reaction $R_0 \to R_2$ will drive product accumulation until $\frac{R^*_2}{R^*_0} \approx e^{0.5/RT}$. Since $e^{0.7/RT}>e^{0.5/RT}$, the former species can invade and grow in the steady-state environment of the latter, driving the product accumulation to levels where the latter cannot grow. Thus the species utilizing the path dissipating more heat is able to invade and displace its competitor, and create conditions that cannot be invaded by the other. Hence, the path to $ R_2$ that dissipates more heat is realized in the final community~(Fig.~\ref{fig:max_heat_dissip}D).

The same argument can be extend to resources with multiple paths leading to them, such as $R_3$. We consider three paths to $R_3$ in Fig.~\ref{fig:max_heat_dissip}E. Note that we exclude the path $R_0\to R_2\to R_3$ since we previously found the maximally dissipative path to $R_2$; this is an application of Prim's algorithm to find the maximal spanning tree in a network~\cite{prim_shortest_1957}. Comparing the heat dissipated among the three paths, we find the maximally dissipative path, $R_0\to R_1\to R_3$. A species utilizing this path is able to raise the concentration of $R_3$ high enough so as to make the other reactions infeasible, driving competitors relying on the paths extinct. Thus the principle of maximum dissipation governs the selection of the community metabolic network in slow-growing, energy-limited communities.

\subsection*{Functional convergence from thermodynamic principles}

The principle of maximum heat dissipation explains why the same reactions are realized in the final communities across pools. But it does not explain the convergence in reaction fluxes across pools. 

To understand how this convergence in fluxes arises, we build on results in Eq.~\eqref{eq:RiRj_at_SS} to derive the concentration at steady state of $R_0$,
\begin{equation}
R_0^*= \frac{h_0}{1+ \sum_{ j=1 }^{M} e^{Q_{0\Rightarrow j }/RT  } }  \;  + \mathcal{O}\left(\frac{\delta}{g_{\mathrm{max}}}\right),
\label{eq:R0_ss}
\end{equation}
where $h_0$ is the steady state concentration of $R_0$ in absence of any consumption and $Q_{0\Rightarrow j }$ is the heat dissipated in the reaction path to $R_j$ realized in the community (see SI). Using Eq.~\eqref{eq:RiRj_at_SS}, we can obtain the the steady-state concentrations of all resources:
\begin{equation}
R_i^*= \frac{h_0}{1+\sum_{ j=1 }^{M} e^{Q_{0\Rightarrow j }/RT  } } e^{ Q_{0\Rightarrow i } /RT}  + \mathcal{O}\left(\frac{\delta}{g_{\mathrm{max}}}\right).
\label{eq:Ri_ss}
\end{equation}
These equations imply that the resource concentrations at steady-state are determined to leading order by thermodynamics alone. The surviving species grow to abundances that allow them to maintain the environmental resource concentrations at these  thermodynamically-determined values. Note that these equations resemble the Boltzmann-Gibbs distribution in statistical physics, now realized in an ecological context.

The fluxes in the steady-state community can be calculated by conserving resource flux along the chemical reaction network. The outflow from a terminal resource without any outgoing reactions, $R_T$, is solely determined by dilution and equal to $\delta R^*_T$. At steady state, this outflow is balanced by the reaction flux into the terminal resource. This can be used to calculate the flow from resources producing a terminal resource. Iterating this procedure, we can calculate the total flux into any resource $R_i$, $I_i$, as 
\begin{equation}
I_i^*=  R_i^* \delta  +  \sum_{ j  \;\text{downstream of}\;  i }  R_j^* \delta  + \mathcal{O}\left(\frac{\delta}{g_{\mathrm{max}}}\right),
\label{eq:flux_fba}
\end{equation}
where the sum is over all resources, $R_j$, that appear downstream of $R_i$ in the reaction network. This theoretical prediction matches simulations (Fig.~\ref{fig:functional_convergence} A). Importantly, the steady-state resource concentrations and reaction fluxes are independent of enzyme budgets, enzyme allocation, and details of the chemical kinetics---none of the associated parameters~($E, k_{cat},  K_M, K_S$) appear in the equations to leading order (Eq.\ref{eq:Ri_ss}). Hence there is strong selection for choosing the maximally dissipative network and converging to the thermodynamically constrained fluxes; selection on the species-specific properties is weaker. Thus, we observe functional convergence, driven by the shared reaction thermodynamics across experiments, despite taxonomic divergence, in terms of how species conspire to maintain the community function.

Note that a community utilizing the maximally dissipative reactions can still experience species turnover.  For example, an invading species that allocates more enzyme to the maximally dissipative reactions can displace a resident. However, this invasion will have only a very small effect (of order $\frac{\delta}{g_{\mathrm{max}}}$) on the community metabolic function and steady-state resource environment. Hence communities can continuously experience species turnover while maintaining their metabolic functions, as seen in bioreactors~\cite{fernandez_how_1999,peces_deterministic_2018}.

\section*{Discussion}

Microbial community models have successfult exposed the organizational principles and universal, emergent behavior of microbial communities structured by nutrient limitation~\cite{macarthur_species_1970, posfai_metabolic_2017,tikhonov_collective_2017, goldford_emergent_2018,marsland_minimum_2020, dubinkina_multistability_2019, wang_complementary_2021}. However, in many natural and industrial environments, microbial communities grow under energy-limited conditions and are forced to use low free energy reactions to drive their growth.\cite{jessen_hypoxia_2017, jorgensen_feast_2007, leng_review_2018, hoh_experimental_1997, nobu_catabolism_2020, bradley_widespread_2020, conrad_thermodynamics_1986, bradley_widespread_2020, hickey_thermodynamics_1991, jackson_anaerobic_2002,larowe_energetics_2019,larowe_thermodynamic_2012}. Here, by explicitly modeling the thermodynamic inhibition of microbial catabolism and energy-limitation of growth, we have developed a minimal model of microbial communities in low-energy environments. We used the model to investigate the emergent properties of slow-growing microbial communities assembled in low-energy environments. Strikingly, model simulations displayed the `functional convergence despite taxonomic divergence' observed in many communities~\cite{fernandez_how_1999, louca_decoupling_2016,louca_function_2018}. We demonstrated that functional convergence in our model originates from the fundamental thermodynamic constraints on microbial growth; previous approaches have shown functional convergence in models of nutrient-limited communities due to shared metabolic constraints between species~\cite{fant_eco-evolutionary_2021}. Further, we demonstrated that community metabolic network structure in slow-growing communities assembled by ecological competition over long periods of time is determined by the thermodynamic principle of maximum dissipation. Importantly, network selection by maximum dissipation manifests through ecological competition and is not expected in isolated microbes.

For simplicity, we demonstrated our main results using a minimal model of energy-limited communities. In the supplementary text, we show that our results hold more generally. In particular, we demonstrate that our results apply in the following scenarios: First, if reaction stoichiometry varies beyond $1:1$ in the model~(see SI Sec.6). Second, if maintenance energy requirements of the cell are incorporated in the model; the maintenance energy flux requirement resembles biomass dilution, and so our results apply as long as this requirement is small~(see SI Sec.5). And third, a model without explicit dilution of resources, where cells uptake resources for anabolic processes in addition to catabolism~(see SI Sec.7).

\begin{figure*}
\includegraphics[width=\textwidth]{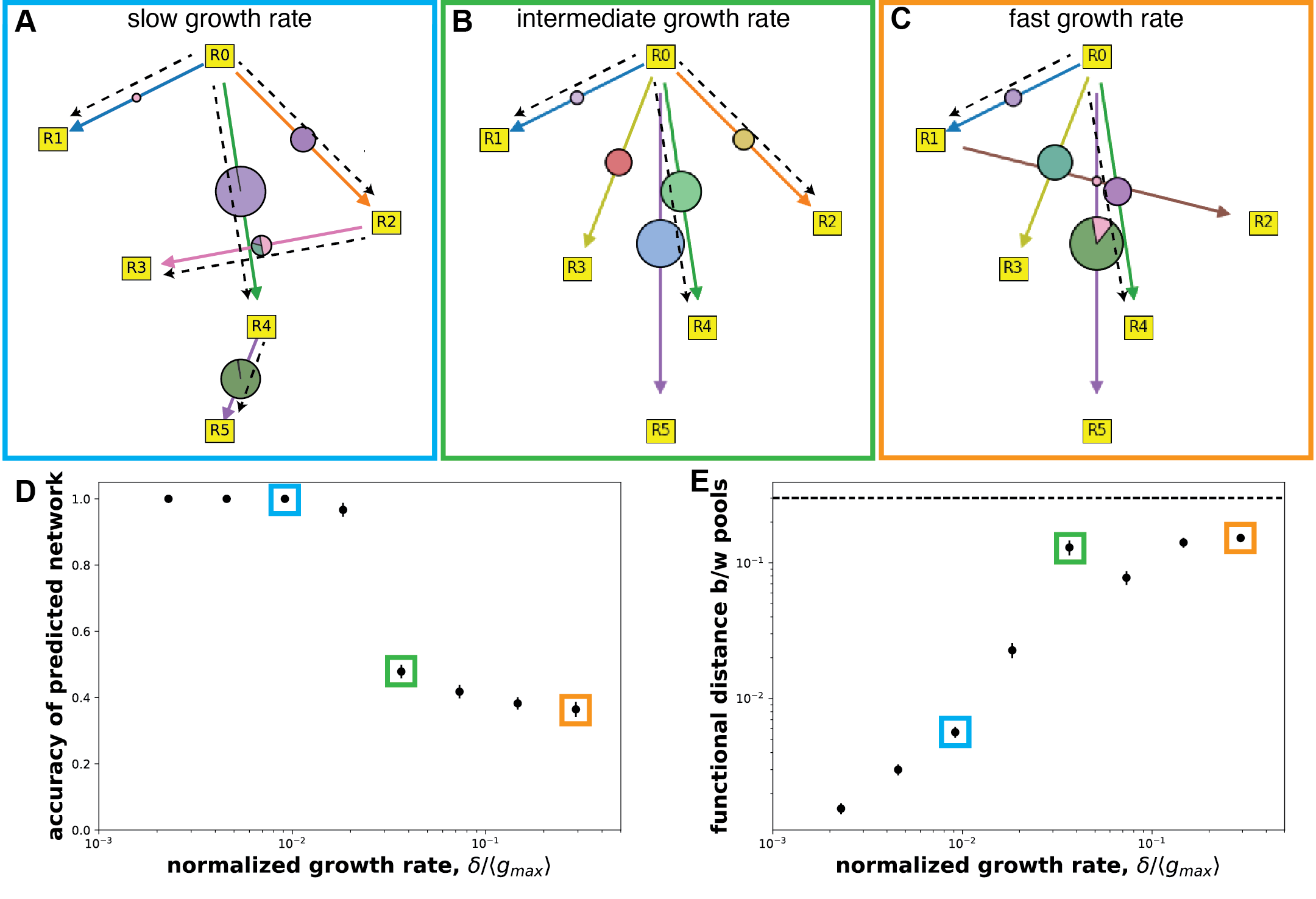}
\caption{\textbf{Functional convergence and the strength of thermodynamic constraints decreases in fast growing communities.} \textbf{A),B),C)} The realized community metabolic networks at slow growth rate, intermediate growth rate, and fast growth rate respectively. All three communities were obtained from the same species pool, and the steady-state growth rate was varied by changing the dilution rate, $\delta$. The dashed lines indicate that the active reaction in the community agrees with the prediction based on maximum dissipation principle. In the slow-growing community, the active reactions match the five reactions predicted from the maximum dissipation principle; in the middle community 3 of the active reactions match the prediction; in the fast-growing community only 2 of the active reactions match the prediction. 
\textbf{D)} The fraction of reactions correctly predicted by the maximum dissipation principle, averaged across 10 different species pools in the same environment. The accuracy decreases as the growth rate increases.
\textbf{E)} The functional distance between two communities was quantified by the Jensen-Shannon distance between the two community flux vectors. The average functional distance increases with the growth rate. Thus, the observed functional convergence decreases as the communities grow faster. The dashed line indicates the average distance between two random flux vectors. A random flux vector had the same length (15), and had each element picked from a uniform distribution between 0 and 1. The colored boxes in panels D and E correspond to the growth rates shown in the panels in A,B,C.
The maximum growth rate, $g_{max}$, was used to normalize the steady-state growth rate. $g_{max}$ was defined as the growth rate obtained on an irreversible reaction between consecutive resources, without any energy lost as heat, by a species investing its entire budget in the reaction (see Methods). 
}
\label{fig:varying_delta}
\end{figure*}

An important assumption in our theoretical derivation of the maximal dissipation principle and thermodynamic explanation of functional convergence was that communities were slow-growing. But what happens when growth rate increases? Our theoretical analysis predicts that differences in species' enzyme budgets and allocation, and in enzymatic parameters, will start to matter as growth rate increase, and will cause greater differences in the community metabolic network structure and function of communities from different pools. To study community metabolic network structure and function as the steady-state growth rate increases, we simulated community assembly from 10 separate pools of 600 species at different dilution rates $\delta$. Since the growth rate of a community at steady-state is equal to the dilution rate, these simulations provided us with 10 different final communities at each steady-state growth rate. 

Fig.~\ref{fig:varying_delta}A,B,C depicts three final communities obtained, at three different dilution rates from the same species pool. The maximum dissipation principle predicts all 5 reactions correctly in the slow-growing community, which grows at $1\%$ of the maximum possible growth rate (dashed lines indicate reactions predicted correctly). At an intermediate growth rate, the maximum dissipation principle predicted 3 out of the 5 reactions correctly. Finally, for the fast-growing community, which grew at $30\%$ of the maximum possible growth rate, 2 out of 5 reactions were predicted correctly. The accuracy of the predictions by the maximum dissipation principle, defined as the average fraction of reactions predicted correctly across the 10 species pools, is shown in Fig.~\ref{fig:varying_delta}D. Communities deviate from the predictions at higher dilution rates due to a combination of factors. First, due to the low energy yields of reactions on the maximally dissipative path, species utilizing these reactions need to catalyze a larger flux to achieve high growth rates than species utilizing reactions with higher energy yields on competing paths. Second, due to the variation in enzyme budgets and enzyme allocation strategies, a species utilizing reactions on a maximally dissipative path can get out-competed by a species with a larger enzyme budget allocated to reactions on a competing path. Third, due to the variation in enzymatic parameters, a reaction on a competing path might be preferred because it has better enzymes, (e.g. with higher $k_{\mathrm{cat}}$). 

As the idiosyncratic properties of species pools and enzymes start to play a bigger role in determining the community metabolic network, functional convergence weakens. Fig.~\ref{fig:varying_delta}E shows that the average distance between community fluxes, or functional divergence, increases in fast-growing communities. The dashed line shows the expected distance between two random flux vectors; this distance is higher than the most functionally distant communities because the random flux vectors are not subject to ecological and flux constraints. Together, the observations in Fig.~\ref{fig:varying_delta} show that the maximum dissipation principle and functional convergence will be stronger in low-energy communities that grow slowly. Intriguingly some published results hint at this loss of functional convergence with faster growth. The functional stability of communities in anaerobic bioreactors was analyzed in three different studies~\cite{fernandez_how_1999,peces_deterministic_2018,zhou_stochastic_2013}, operating at dilution times of 10 days, 15 days and 1 day. While the two studies at the smaller dilution rates observed functional convergence, the study at the highest dilution rate did not.

Microbes in low-energy environments face a trade-off between harvesting energy efficiently, to increase biomass yield per unit reaction flux, and dissipating more heat, to increase reaction flux. A microbe cannot grow at either extreme---assimilating all the energy curbs growth because the reaction flux will be zero, while dissipating all the energy curbs growth because energy assimilated will be zero. Thus, microbes have to find a balance between these opposing forces. Our results suggest that, at small dilution rates, the effect of ecological competition will push microbes to evolve towards dissipating more energy. However, in practice this evolutionary drive to decrease energy assimilation will not continue forever. It will probably be limited by biochemical constraints~\cite{schink_syntrophism_2006,schink_energetics_1997, lever_life_2015,hoehler_biological_2004}, such as the smallest energy unit that can be assimilated by a species (one electron, one ATP, etc.), or when reactions become limited by substrate depletion rather than thermodynamic inhibition. In our simulations, to focus on the ecological properties of energy-limited communities, a species at the lower-limit of energy assimilation could survive at the smallest dilution rates considered. An exploration of the evolutionary consequences of the selective pressures on energy assimilation is reserved for future work.

Optimization principles grounded in thermodynamics have been used in microbial metabolism to explain behavior at the cellular scale~\cite{henry_thermodynamics-based_2007,flamholz_glycolytic_2013, goldford_protein_2021}. At the community level, optimization of thermodynamic quantities has been conjectured to be a mechanism by which communities self-organize, motivated mostly by verbal arguments and analogies with physical systems~\cite{schneider_life_1994, vallino_ecosystem_2010}. However, the question of which quantity should be maximized, and under what conditions, remains actively disputed in the literature~\cite{dewar_theoretical_2014, meysman_ecosystem_2010}. By incorporating ecological and thermodynamic principles, our model connects these two disparate approaches toward understanding microbial communities. We derive the thermodynamic optimization principle of maximum dissipation, which structures slow-growing, energy-limited communities, providing a concrete example of thermodynamic optimization in ecology.

Our results make predictions for the metabolic structure and resource environment created by the communities in low-energy environments, based on the thermodynamics of microbial metabolism. This requires knowledge of the free energy differences and metabolomic data of the resource environment. Although calculating the free energies of the many metabolites present in a real environment is challenging, recent advances in computational methods make the outlook promising~\cite{noor_consistent_2013,jinich_quantum_2018, jinich_mixed_2019}. Combining these methods with metabolomic data and knowledge of microbial metabolism to predict metabolite composition in real environments shaped by microbial communities is an exciting prospect for future research.\\

\textbf{Acknowledgements.} The authors thank Avi Flamholz for valuable comments and feedback on the manuscript. This research was supported in part by NSF Grant No. PHY-1748958 and the Gordon and Betty Moore Foundation Grant No. 2919.02.

\section*{Methods}

\subsection*{Species and resource dynamics}
Species dynamics are determined the amount of reaction flux they can utilize, the energy the assimilate from this to create biomass, and biomass loss due to dilution. The dynamics of species $\alpha$ with abundance $N^{(\alpha)}$ is described by
\begin{equation}
\frac{d N_{\alpha}}{d t}= N_{\alpha} \left[ \sum_{i,j \in \mathcal{X} } E^{(\alpha)}_{i j} \mathcal{F}_{ij} \mathcal{E}^{\mathrm{ATP}}_{ij} Y  -\delta \right],
\label{eq:dNdt}
\end{equation}
where $F_{i j}$ is the flux of reaction $R_i \to R_j$ catalyzed per unit enzyme, $E^{(\alpha)}_{i j}$ is the enzyme allocated to reaction $R_i \to R_j$ by species $\alpha$, $\mathcal{X}$ is the set of all allowed reactions, and $\delta$ is the dilution rate. The maintenance energy requirements of a species can be accounted for by replacing $\delta$ with $\delta +m$ in Eq.~\ref{eq:dNdt}, where $m$ is the energy flux required for cell maintenance. Our results apply in this scenario as well (see SI).

Resource dynamics are determined by their consumption and production via the reactions catalyzed by the species, external supply, and dilution. The dynamics of resource $i$ with concentration $R_i$  are given by
\begin{equation}
\frac{d R_i}{d t}= h(R_i) - \sum_{\alpha, j} N_{\alpha} E^{(\alpha)}_{i j} \mathcal{F}_{ij}+ \sum_{\alpha, k} N_{\alpha} E^{(\alpha)}_{ki} \mathcal{F}_{ki} - \delta R_i
\label{eq:dRdt}
\end{equation}
where $h(R_i)$ is the resource supply. Only the top resource $R_0$ is supplied, so $h(R_0)=\delta h_0$ and zero otherwise. $h_0$ is the concentration of resource in the supply.

\subsection*{Simulation procedure}

All species in a pool were introduced at the same initial density (=1) in an environment with resource concentration of the supplied resource $R_0=h_0$ and all other resources at a very small concentration $10^{-10}$. The species, reaction, and resource dynamics were simulated till they reached a steady-state. At regular intervals, species that fell below an abundance threshold had their concentration set to zero. The minimum allowed resource concentration was $10^{-10}$ to prevent division by zero in Eq. \eqref{eq:JSP}. A steady-state was reached if the maximum value of the vector $\frac{1}{N_{\alpha}} \frac{d N_{\alpha}}{d t}$, measuring the logarithmic growth rates, fell below $10^{-5}$. 

\subsection*{Simulation parameters}

The enzyme budget of a species was chosen from a lognormal distribution with lognormal parameters $\mu=1, \sigma=0.1$. A species utilized a random subset of the 15 possible reactions. So species in the pool were not biased towards extreme generalists or specialists. The enzyme budget was allocated to the selected reactions uniformly, by using a Dirichlet distribution with the dirichlet parameter, $\alpha=1$, for the selected reactions. There were 600 species in each of the 5 pools in Fig.\ref{fig:functional_convergence} and 10 pools in Fig.~\ref{fig:varying_delta}; species were not shared across pools.

The energies of resources $R_0, R_1,R_2, R_3, R_4,$ and $R_5$ were $5RT, 4RT, 3RT, 2RT, 1 RT,$ and $0RT$. For each reaction, a random fraction between $15\%$ and $85\%$ of the standard energy gap between product and substrate ($\mathcal{E}_S- \mathcal{E}_P$) was assimilated and the rest was dissipated as heat. This assimilated energy, along with resource energies, was fixed across the experiments with different species pools. The concentration of $R_0$ in the supply, $h_0=80$. The chemical kinetic parameters $K_P$,$K_S$, and $k_{cat}$ were chosen from lognormal distributions with lognormal parameters $\mu=0., \sigma =0.01$, and so, their mean was close to 1. The energy to biomass yield factor,~$Y=1$. The dilution rate,~$\delta=0.01$ in Fig.~\ref{fig:functional_convergence}; this rate was small enough that a typical species would survive if it was alone.

The Jensen-Shannon distance, used to measure functional distance between two flux vectors in Fig.~\ref{fig:varying_delta}. It is the symmetrized analog of the Kullback-Leibler divergence, and was defined as 
\begin{equation}
\mathrm{JSD} (f^{(a)}, f^{(b)}) = \frac{1}{2} \sum_i f^{(a)} \log \frac{f^{(a)}}{f^{(m)}}  + \frac{1}{2} \sum_i f^{(b)} \log \frac{f^{(b)}}{f^{(m)}} ,
\end{equation}
where $f^{(a)}, f^{(b)}$ are the two normalized flux vectors and $f^{(m)}$ is an average flux vector defined as $(f^{(a)}+ f^{(b)})/2$. We report the JSD averaged over all pairs of communities in Fig.~\ref{fig:varying_delta}.

$g_{max}$ in Fig.~\ref{fig:varying_delta} was defined as the growth rate obtained from an irreversible reaction where a species invested its entire enzyme budget into one reaction. Since this varies between reactions because reactions different energy yields, we chose the energy gap between consecutive resources. It was calculated as the product of mean $k_{cat}$, mean enzyme budget, energy gap between consecutive resources ($=RT$), and biomass yield factor $Y$.

\bibliography{bib_thermo}
\bibliographystyle{naturemag}

\end{document}


\title{Supplementary text\\Functional universality in slow-growing microbial communities arises from thermodynamic constraints}
\author{Ashish B. George, Tong Wang, and Sergei Maslov }
\date{\today}
\maketitle
\tableofcontents

\section{Model}

We consider a community model where species grow by harvesting energy from chemical reactions converting metabolic substrates into products. There are $M+1$ resources or metabolites that we label in the order of decreasing energy as $R_0,R_1,...,R_M$, with energies $\mathcal{E}_0,\mathcal{E}_1,...,\mathcal{E}_M$ such that $\mathcal{E}_0>\mathcal{E}_1>...\mathcal{E}_M$. Species utilize reactions that convert a substrate of a higher energy into a product of lower energy. This mirrors the trend in catabolism of degrading more complex and energy-rich molecules, like glucose into simpler ones, like acetate.

Chemical reactions connect all pairs of resources, making the space of possible reactions a fully connected network with $(M+1)M/2$ possible links. From each unit flux of reaction $R_i\to R_j$ , species harvest $\mathcal{E}^{\mathrm{ATP}}_{ij}$ of energy, stored in ATP or other forms. The species convert harvested energy into biomass based on a yield factor $Y$ which is constant across species. For a species to utilize a reaction, it needs to have the corresponding enzyme. Each species allocates an overall enzyme budget $E^{\mathrm{total}}$ among reactions it can catalyze. We allow overall enzyme budgets to vary between species to avoid any special degenerate behavior~\cite{tikhonov_collective_2017}. 

{Reversible Michaelis-Menten kinetics describes reaction flux}
We illustrate the scheme to calculate flux through the chemical reactions by using an example reaction below.  A enzyme catalyzed reaction from substrate $S$ to product $P$ can be understood as following the scheme:
\begin{equation}
E +S  \rightleftharpoons  E X\rightleftharpoons E +P,
\end{equation}
where $E$ is the enzyme corresponding to this reaction and $EX$ is some intermediate enzyme complex. The conversion of the substrate to the product could proceed over multiple steps; this will not make a difference to our results~\cite{jin_thermodynamics_2007}. One can also think of this reaction as a coarse-grained or lumped description of the full reaction ~\cite{wachtel_thermodynamically_2018, seep_reaction_2021,henry_thermodynamics-based_2007}.

The cell harvests energy, $\mathcal{E}^{\mathrm{ATP}}$, from the reaction, which is stored in the phosphorylation of ADP, reduction of NAD, or other forms. The energy harvesting decreases the Gibbs free energy drop across the reaction. The free energy difference under standard conditions is
\begin{equation}
\Delta G^0= \mathcal{E}_P^0 + \mathcal{E}^{\mathrm{ATP}}- \mathcal{E}_S^0,
\end{equation}
where $\mathcal{E}_P, \mathcal{E}_S$ are the standard-state energies of the substrate and product. The energy dissipated as heat, $Q$ is equal to negative of this free energy difference, i.e, $Q=-\Delta G^0$.

The reaction flux is described by reversible Michaelis-Menten kinetics:  
\begin{equation}
J= k_{cat} E  \frac{ \frac{S}{K_S}   }{1 +  \frac{S}{K_S} + \frac{P}{K_P}  } \left(1- \frac{P}{S} \gamma \right),
\label{eq:J}
\end{equation}
where $k_{cat},K_S, K_P$ are enzymatic parameters describing chemical kinetics and $ \gamma= e^{- Q /RT}$.  The first term, $k_{cat} E$, describes the maximal velocity of the reaction, which we call $\nu$. The second term describes enzyme binding affinities and saturation level. The final term describes inhibition due to product accumulation. If $P$ is very large, the equation predicts a negative flux due to the reversal of the reaction direction. This would cause species to lose energy from using the reaction. So we assume that species down-regulate enzyme production to prevent energy loss from driving the reverse reaction.

We will also define the reaction flux per unit enzyme, $\mathcal{F}$ as 
\begin{equation}
\mathcal{F}=J/E= k_{cat}   \frac{ \frac{S}{K_S}   }{1 +  \frac{S}{K_S} + \frac{P}{K_P}  } \left(1- \frac{P}{S} \gamma \right),
\label{eq:flux_per_E}
\end{equation}

\subsubsection*{The low and high substrate flux limits}

The low and high flux regimes correspond to $S^* \ll K_S$ and $S^* \gg K_S$ respectively (see~\cite{noor_note_2013} for a detailed discussion of the limits).

In the low flux regime, the consumption flux is proportional to substrate concentration, i.e.,
\begin{equation}
J=k_{cat} E  \frac{S}{K_S}\left(1- \frac{P}{S} \gamma \right).
\label{eq:lowflux}
\end{equation}

In the high flux regime $S^*/K_S \gg 1 + P^* / K_P$, the consumption flux is substrate saturated
\begin{equation}
J=k_{cat} E \left(1- \frac{P}{S} \gamma \right).
\label{eq:highflux}
\end{equation}

\subsubsection*{Extending notation to multiple reactions}

The flux equations described above were for a single reaction. To apply these equations in the model, which has many reactions, we add subscripts to denote the reaction. The enzyme allocated by an individual of species $\alpha$ for a reaction $R_i \to R_j$ will be $E^{(\alpha)}_{ij}$. The corresponding per-capita flux will be $J^{(\alpha)}_{ij}$. The chemical kinetic constants and heat dissipated are chemical and enzymatic properties which vary only between reactions and not  between species. Thus the flux per unit enzyme is $\mathcal{F}_{ij}$, the heat dissipated is $Q_{ij}$, the strength of inhibition is $\gamma_{ij}$, etc.  Since the notation becomes clumsy quickly, we will suppress indices for clarity if possible.

\subsection*{Species and resource dynamics}

Species dynamics are determined the amount of reaction flux they can utilize, the energy the assimilate from this to create biomass, and biomass loss due to dilution. The dynamics of species $\alpha$ with abundance $N^{(\alpha)}$ is described by
\begin{equation}
\frac{d N_{\alpha}}{d t}= N_{\alpha} \left[ \sum_{i,j \in \mathcal{X} } E^{(\alpha)}_{i j} \mathcal{F}_{ij} \mathcal{E}^{\mathrm{ATP}}_{ij} Y  -\delta \right],
\label{eq:dNdt}
\end{equation}

where  $F_{i j}$ is the flux of reaction $R_i \to R_j$ catalyzed per enzyme, $E^{(\alpha)}_{i j}$ is the enzyme allocated to reaction $R_i \to R_j$  by species $\alpha$, $\mathcal{X}$ is the set of all allowed reactions, and $\delta$ is the dilution rate. The flux in a particular reaction catalyzed by a species is zero if the species does not allocate any enzyme to it. 

Resource dynamics are determined by their consumption and production via the reactions catalyzed by the species, external supply, and dilution. The dynamics of resource $i$  with concentration $R_i$  are given by
\begin{equation}
\frac{d R_i}{d t}= h(R_i) - \sum_{\alpha, j} N_{\alpha} E^{(\alpha)}_{i j} \mathcal{F}_{ij}+ \sum_{\alpha, k} N_{\alpha} E^{(\alpha)}_{ki} \mathcal{F}_{ki} - \delta R_i
\label{eq:dRdt}
\end{equation}
where $h(R_i)$ is the resource supply. The second and third terms described consumption and production fluxes of the resource by the species, with $j$ indexing the resources that could be produced from $i$ and $k$ indexing the resources that could produce $i$.

We study the scenario where the most energetic resource, $R_0$ is supplied. The concentration of $R_0$ in the supply is $h_0$. Thus the supply function is
\begin{equation}
  h(R_i) =
  \begin{cases}
    h_0 \delta & \text{if $i = 0$ } \\
    0 & \text{otherwise}.
  \end{cases}
\end{equation}

\section{Solution for the steady-state community}
In this section, we will examine the equations describing a given community at steady state. At steady state, we can set the LHS of Eq.\eqref{eq:dNdt} to zero for any surviving species $\alpha$. This gives us the condition
\begin{equation}
\sum_{i,j \in \mathcal{X}} E^{(\alpha)}_{i j} \mathcal{F}_{ij} \mathcal{E}^{\mathrm{ATP}}_{ij} Y  = \delta \; \forall \alpha \in \mathrm{survivor} .
\label{eq:no_matrix_ss}
\end{equation}
By counting the number of independent equations and degrees of freedom, we can see that the number of surviving species is limited by the number of reactions. Furthermore, because species regulate enzymes to ensure they do not catalyze reaction fluxes in the reverse direction and lose energy, the number of survivors is limited by the number of active reactions (reactions that carry a positive flux).

Restricting the equations to just the surviving species and active reactions only, we can express the Eq.~\eqref{eq:no_matrix_ss} as a matrix equation:
\begin{equation}
 \mathbf{E}  \mathcal{\overrightarrow{F}} Y =\delta \overrightarrow{1},
\end{equation}
where $ \mathbf{E}$ is the matrix of enzyme allocation of the surviving species in the active reactions, $\mathcal{\overrightarrow{F}}$ is the vector of fluxes in the active reactions, $\delta$ is the dilution rate, and $\overrightarrow{1}$ is a vector of ones corresponding to each survivor.

We can solve this equation by multiplying the inverse of $\mathbf{E}$ matrix on both sides. Substituting the definition of $\mathcal{F}$ (Eq. \eqref{eq:flux_per_E}), we get
\begin{equation}
\left(1-\gamma_{ij} \frac{R^*_j}{R^*_i}\right)  =\frac{\delta}{k_{cat ,ij} \mathcal{E}^{\mathrm{ATP}}_{ij} Y }   \frac{1+\frac{R^*_i}{K_{S_{ij}}} +\frac{R^*_j}{K_{P_{ij}}  }  }{ \frac{R^*_i}{K_{S_{ij}}}   } \left[\mathbf{E^{-1}}  \overrightarrow{1}\right]_{ij} ,
\label{eq:matrix_Inv}
\end{equation}
for each active reaction $R_i \to R_j$ in the community, where $*$ is used to denote the steady state concentration of the corresponding resource and $\left[\mathbf{E^{-1}}  \overrightarrow{1}\right]_{ij} $ refers to the element of the vector $\mathbf{E^{-1}}  \overrightarrow{1}$ corresponding to reaction $R_i \to R_j$. 

We can further simplify this in the high-flux limit (Eq. \eqref{eq:highflux}) where the second term in the RHS is one. Physically, this means that we are assuming that reaction fluxes are primarily slowed down by thermodynamic inhibition rather than substrate depletion. For each active reaction in the final community, we have a relation for the steady state concentrations of the substrate and product:
\begin{equation}
\frac{R^*_j}{R^*_i}= \frac{1}{\gamma_{ij}}  \left( 1-   \frac{\delta}{k_{cat ,ij} \mathcal{E}^{\mathrm{ATP}}_{ij} Y }  \left[\mathbf{E^{-1}}  \overrightarrow{1}\right]_{ij} \right).
\label{eq:RjRi_ss_matrix}
\end{equation}
Thus $\frac{R^*_j}{R^*_i}$ is set by thermodynamics to leading order in $\delta$. 

To simplify the notation, we will define $\mu_{ij}$ as
\begin{equation}
\mu_{ij}=  \frac{\delta}{k_{cat ,ij} \mathcal{E}^{\mathrm{ATP}}_{ij} Y }  \left[\mathbf{E^{-1}}.  \overrightarrow{1}\right]_{ij} 
\end{equation}

To gain intuition we consider the expression when species are specialists. In this case we have:
\begin{equation}
\mu_{ij}=  \frac{\delta }{k_{cat ,ij} \mathcal{E}^{\mathrm{ATP}}_{ij} Y E^{(\beta_{ij})}_{ij}  } ,
\end{equation}
where $\beta_{ij}$ is the species specializing in reaction $R_i \to R_j$. We see that $\mu_{ij}$ compares the steady-state growth rate of the species in the community ($\delta$) to the maximum growth rate of the species, achieved in an environment without thermodynamic inhibition. Thus $\mu$ is a measure of the degree of thermodynamic inhibition of the species in the community. The specialist approximation can also be of practical use for obtaining rough estimates since we need not have access to the enzyme allocation matrix of a natural community.

In terms of $\mu_{ij}$, the steady-state concentration of the substrate and product in an active reaction is given by
\begin{equation}
\frac{R^*_j}{R^*_i}= \frac{1}{\gamma_{ij}}  \left( 1-   \mu_{ij}\right).
\label{eq:RjRi_ss_matrix_a}
\end{equation}

In addition to the above, the steady state equations also give us an additional condition:
\begin{equation}
R_0^* = h_0- \sum_i R_i^*.
\label{eq:R0_fba}
\end{equation}
This is equivalent to flux conservation, the total inflow of resources must match the total outflow of resources.

Together, we can derive the steady-state concentration of all the resources. We will explain the solution for a general network after illustrating the solution for two special limiting network scenarios.

\subsection*{Hub and spokes network}
\begin{figure}
\includegraphics[width=.5\linewidth]{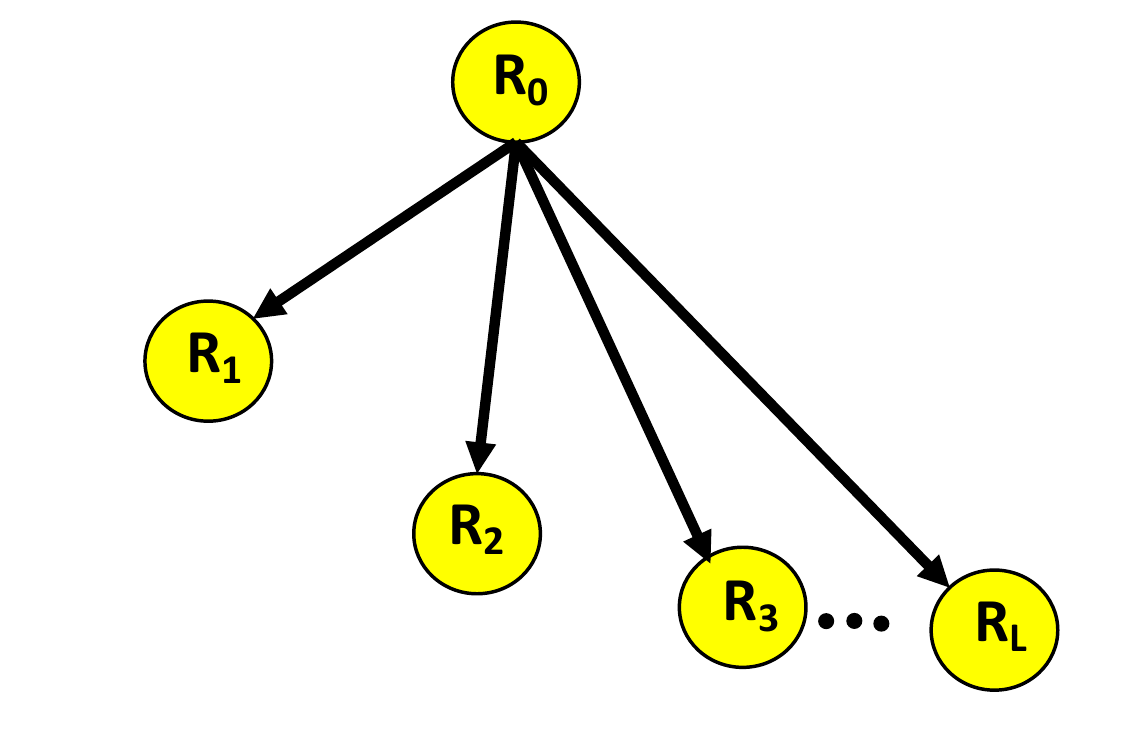}
\caption{\textbf{The reactions in the hub and spokes network.} All reactions start from $R_0$. } 
\label{fig:SI_cartoon1} 
\end{figure}
The first network we consider is a hub and spoke network~(Fig.~\ref{fig:SI_cartoon1}). All active reactions emanate from the sole supplied resource $R_0$, giving products $R_1$..$R_M$. We can solve the system of equations we derived above for this network in a straightforward manner. Eq.\eqref{eq:RjRi_ss_matrix_a} reduces to:

\begin{equation}
\frac{R^*_i}{R^*_0}= \frac{1}{\gamma_{0i}}  \left( 1-   \mu_{0i} \right) \; \; \forall i \in \{1...M\}.
\label{eq:RjRi_ss_hub}
\end{equation}

Solving these equations and Eq.\eqref{eq:R0_fba} we get:
\begin{equation}
R_0^*= \frac{h_0}{1+ \sum_{j=1}^L  \frac{1}{\gamma_{0j}} \left( 1-  \mu_{0j} \right) },
\label{eq:R0_ss_hub}
\end{equation}
and 
\begin{equation}
R_i^*= R_0^*  \frac{1}{\gamma_{0i}} \left( 1- \mu_{0i} \right).
\end{equation}

\subsection*{Linear chain network}

\begin{figure}
\includegraphics[width=.5\linewidth]{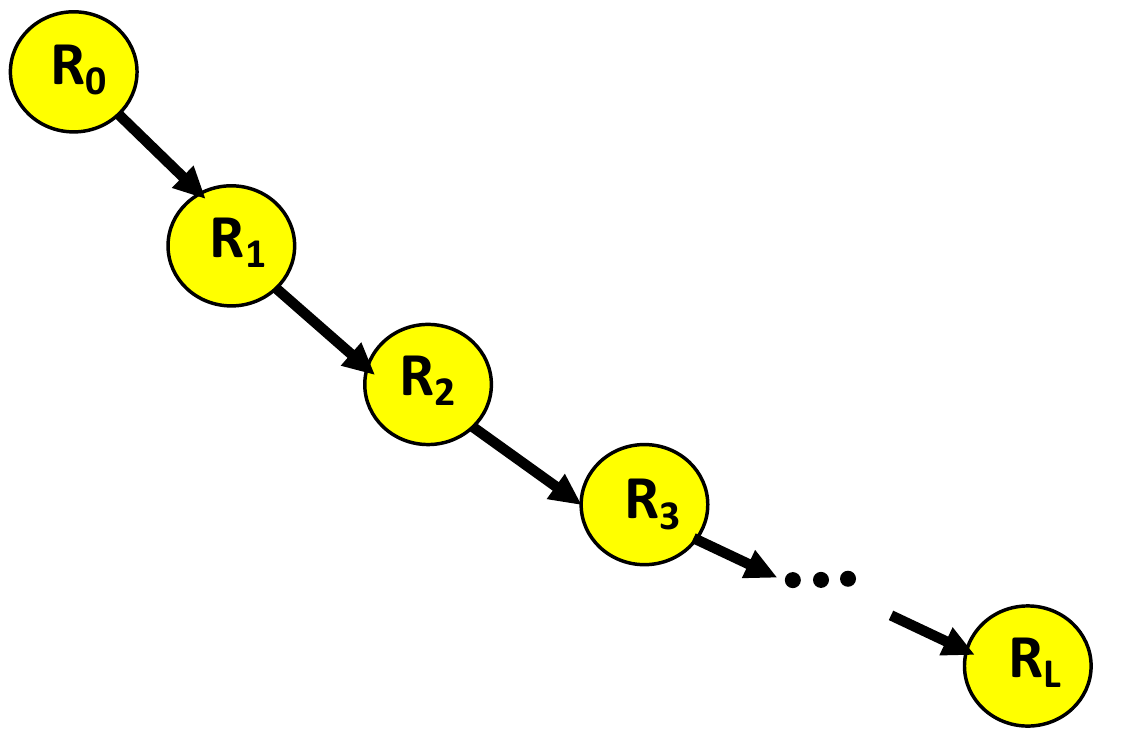}
\caption{\textbf{The linear reaction chain network.}  } 
\label{fig:SI_cartoon2} 
\end{figure}

The second network we consider is the opposite limit, of a linear reaction chain~(Fig.~\ref{fig:SI_cartoon2}). The active reactions connect resources closest in energy. Thus the network forms a chain starting at $R_0$, proceeding down the resource energy hierarchy, and ending at the lowest energy resource $R_M$. Here, Eq.\eqref{eq:RjRi_ss_matrix_a} reduces to:

\begin{equation}
\frac{R^*_i}{R^*_{i-1} }= \frac{1}{\gamma_{i-1,i}}  \left( 1-   \mu_{i-1,i} \right) \; \; \forall i \in \{1...M\}.
\label{eq:RjRi_ss_chain}
\end{equation}

We can recursively solve these equations to express every resource concentration in terms of $R_0^*$, and use Eq.\eqref{eq:R0_fba} to get:
\begin{equation}
R_0^*= \frac{h_0}{1+ \sum_{j=1}^{L} \prod_{k=1}^j \frac{1}{\gamma_{k-1,k}} (1- \mu_{k-1,k}) },
\end{equation}
and
\begin{equation}
R_i^*=R_0^* \prod^i_{k=1} \frac{1}{\gamma_{k-1,k} } \mu_{k-1,k}.
\end{equation}

This expression clarifies that when reactions occur serially, we have a separate product appear for each resource in the chain.

\subsection*{General network}

\begin{figure}
\includegraphics[width=.5\linewidth]{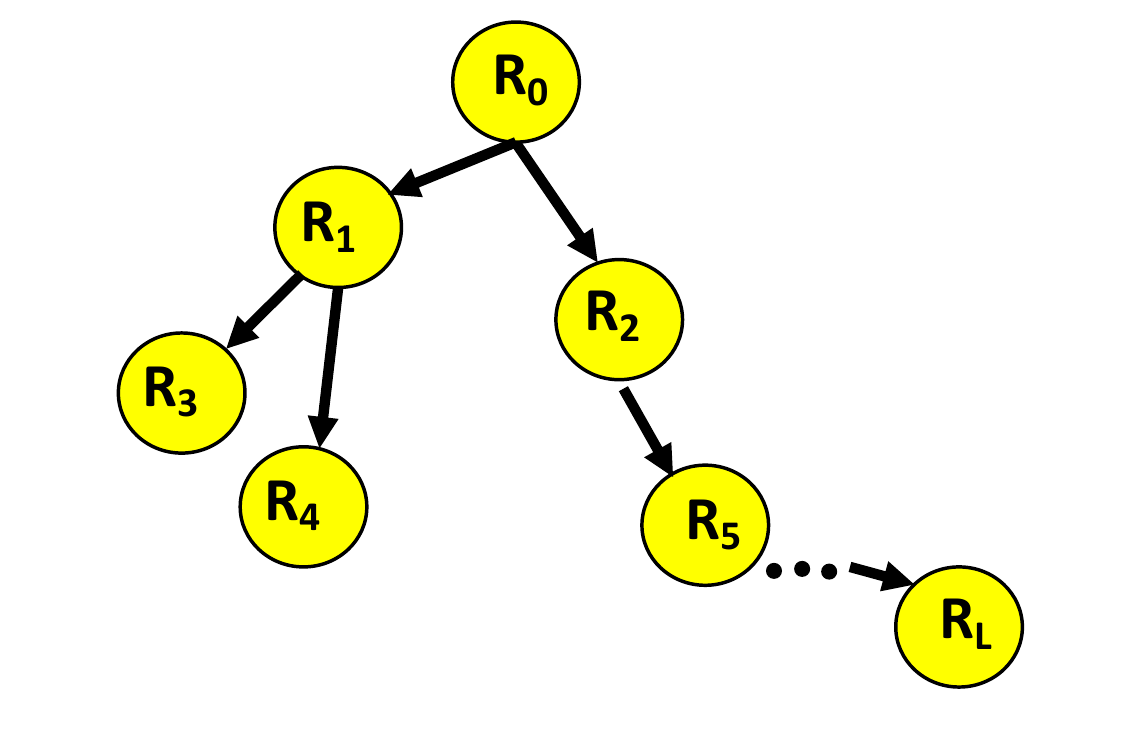}
\caption{\textbf{A general tree network.} Resource nodes can have any number of outflow reactions to lower energy products. No node can have more than one inflow reaction. } 
\label{fig:SI_cartoon3} 
\end{figure}

In the slow-growth limit, the in-degree of a resource is constrained to be 1---thus the network is a tree. We can solve the steady-state equations for an arbitrary tree network using the insights gained from solving the special limits above.

We have the following steady state equations for the tree network:
\begin{equation}
\frac{R^*_j}{R^*_i}= \frac{1}{\gamma_{ij}}  \left( 1-   \mu_{ij} \right) \; \; \forall (i,j) \in \mathcal{X},
\label{eq:RjRi_ss_tree}
\end{equation}
where $\mathcal{X}$ is the set of active reactions in the network. To solve this system, we need to express the steady-state concentration of each resource in terms of the supplied resource, $R_0$. Since the network is a tree, $R_0$ is connected to a resource $R_j$ via a unique path, composed of multiple reactions. We represent the path in short as $R_0 \rightarrow R_j$, and the set of reaction on the path as $\mathcal{P}_{0j}$. Each path resembles the linear chain solved in the previous section. Using this insight, we find the steady-state concentrations as:

\begin{equation}
R_0^*= \frac{h_0}{1+ \sum_{ k=1 }^M \prod_{i,j \in \mathcal{P}_{0\rightarrow k}}  \frac{1}{\gamma_{ij}}  \left( 1-   \mu_{ij} \right) },
\label{eq:R0_tree}
\end{equation}
The pathway to each possible product contributes a term to the sum. To leading order, this term is a product of the heat dissipated on the pathway.

The steady state solution for $R_k$ is 
\begin{equation}
R_k^*= R_0^* \prod_{i,j \in \mathcal{P}_{0k}} \frac{1}{\gamma_{ij}}  \left( 1-   \mu_{ij} \right)
\label{eq:Ri_tree}
\end{equation}

When the steady-state growth of a community is very slow ($\mu_{ij} \ll 1$ ), we have
\begin{equation}
R_k^* \approx R_0^* \prod_{i,j \in \mathcal{P}_{0k}} \frac{1}{\gamma_{ij}}  
\label{eq:Ri_tree2}
\end{equation}

Note that the above analysis assumes that the identities of the surviving species and active reactions are provided. It does not consider competition between species and explain which reactions will be active in the community. We will consider species competition in the following section. 

\section{Selection of metabolic network: principle of maximum dissipation}

In the preceding section, we solved for the steady-state community assuming that we know the active reactions and the surviving species that utilize them. In this section, we will identify the reactions that survive in the community.

\begin{figure}
\includegraphics[width=.5\linewidth]{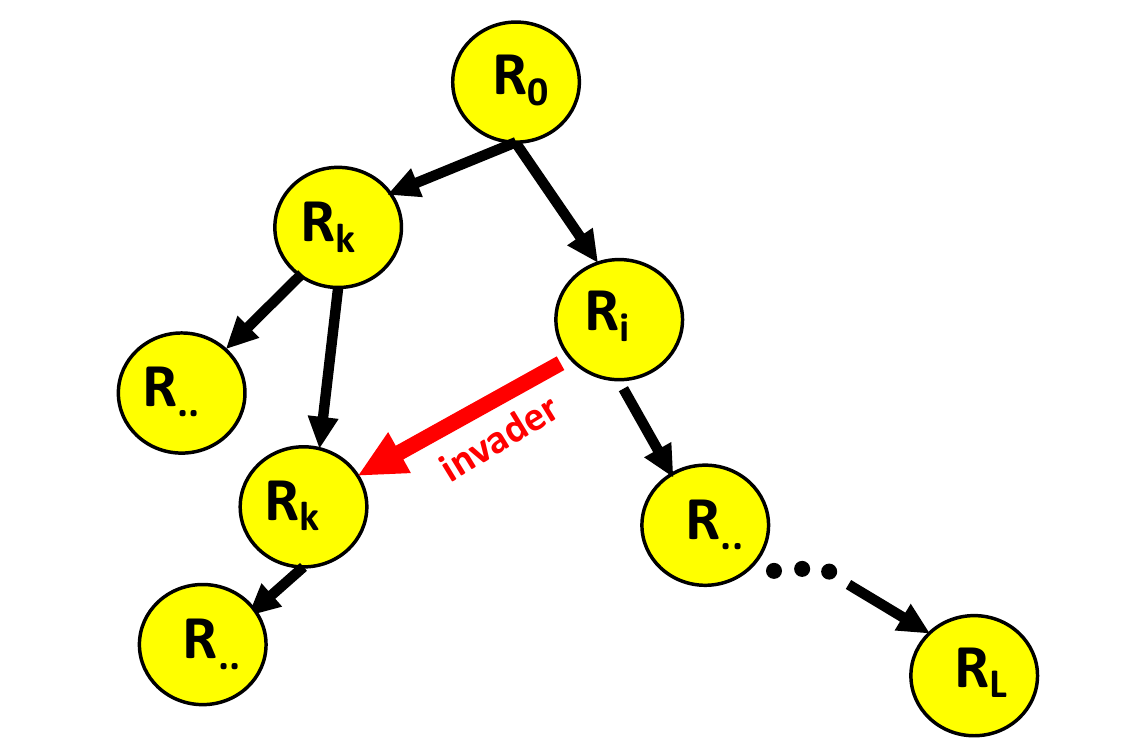}
\caption{\textbf{Analysing invasion of an alternative path reveals network selection principle.} A steady-state metabolic network, depicted by the black arrows, is invaded by a species utilizing the red path from $R_i$ to $R_k$. This path could be composed of multiple reactions. The conditions determining the survival of the invader shows that the network is selected by the maximum dissipation principle.    } 
\label{fig:SI_cartoon_invasion} 
\end{figure}

From an ecological point of view, we can perform an invasion analysis to identify the active reactions. Given a steady state community, we can consider the invasion of  species $\beta_{ik}$ specializing on a reaction $R_i \to R_k$ that is inactive in the current community (Fig.~\ref{fig:SI_cartoon_invasion}). For the invader to succeed, we need it to have a positive growth rate in the environment it is invading, i.e.,
\begin{equation}
k_{cat ,ik}E^{(\beta_{ik})}_{ik} Y\mathcal{E}^{\mathrm{ATP}}_{ik} \left(1- \gamma_{ik} \frac{R_k^* }{R_i^*}\right)  - \delta>0,
\end{equation}
where $R_k^*, R_i^*$ are the steady-state resource concentrations in the community prior to invasion. Using  (Eq.\eqref{eq:Ri_tree}), we can simplify this invasion condition as

\begin{equation}
\frac{\prod_{u,v \in \mathcal{P}_{0k}} \frac{1}{\gamma_{uv}}  \left( 1-   \mu_{uv} \right) }{\prod_{u\prime,v\prime \in \mathcal{P}_{0i}} \frac{1}{\gamma_{u\prime v\prime}}  \left( 1-   \mu_{u\prime v\prime} \right)  } < \frac{1}{\gamma_{ik}}  \left( 1-   \mu_{ik} \right),
\label{eq:max_dissip_mu}
\end{equation}
where the paths $ \mathcal{P}_{0k}, \mathcal{P}_{0i} $ refer to the reaction paths in the community before invasion

We are interested in the slow-growth limit, which is equivalent to small dilution rate $\delta$. For small enough $\delta$ such that $\mu_{ij}= \frac{\delta }{k_{cat ,ij} \mathcal{E}^{\mathrm{ATP}}_{ij} Y E^(\beta)_{ij}  } \ll 1$ for any reaction, the above expression simplifies to 

\begin{equation}
\frac{\prod_{u,v \in \mathcal{P}_{0k}} \frac{1}{\gamma_{uv}}  }{\prod_{u\prime,v\prime \in \mathcal{P}_{0i}} \frac{1}{\gamma_{u\prime v\prime}}   } < \frac{1}{\gamma_{ik}}  
\end{equation}

Using the definition of $\gamma$, this implies that the invader successfully invades if 

\begin{equation}
Q_{0\rightarrow k}< Q_{0\rightarrow i} +Q_{i k} ,
\label{eq:max_dissip_Q}
\end{equation}
where $Q_{0\rightarrow k}$ and $Q_{0\rightarrow i}$ refer to the total heat dissipated over the reactions on the paths $R_0 \rightarrow R_k$ and $R_0 \rightarrow R_i$ in the steady state community before invasion, and $Q_{i k}$ is the heat dissipated in the reaction utilized by the invader. Hence the RHS equals the heat dissipated on the new path from $R_0$ to $R_k$ created by the invader, and the equation is comparing the heat dissipated on the two possible paths to $R_k$.
Thus the invader is able to successfully invade only if the new path to $R_k$ dissipates more heat than the path to $R_k$ in the resident community. 

Over many invasions, the community metabolic network will converge to utilize the paths to each resource that dissipates most heat, and therefore be uninvasible. Thus the principle of maximum dissipation determines the reaction network that is selected by the final community.

Note that the final community can still be successfully invaded by species utilizing the maximally dissipative reactions but happen to be better competitors due to either allocation more enzyme to these reactions or having better enzymes to utilize these reactions. These effects are described in the $\mu$ dependence of Eq.\eqref{eq:max_dissip_mu}, which appears as a lower order contribution for slow-growing communities. Thus communities could continuously experience species turnover while maintaining their metabolic functions.

An alternate way to understand the principle of maximum dissipation is to consider the ratio of $R_k^*/R_0^*$ in the communities differing only by the reaction leading to $R_k$. The community dissipating more heat on the path to $R_k$ sustains a higher value of $R_k^*/R_0^*$. Thus the species utilizing this maximally dissipative path is able to invade the second community successfully but not vice-versa. Since this argument applies to all resources $R_1...R_M$, the final community, that has survived many invasions, utilizes the maximally dissipative path.

\section{Convergence of metabolic fluxes}

The fluxes in the active reactions in the community can be calculated using our results for the steady-state resource concentrations (Eqs.\eqref{eq:R0_tree},\eqref{eq:Ri_tree}) by applying one additional physical principle: conservation. The resource flux via reactions into each resource node in the network is balanced by the flux out of the node via reactions and dilution. The dilution flux of a resource is determined by the dilution rate. Thus the total flux into resource $R_i$, $I_i$, as
\begin{equation}
I_i^*= \delta  R_i^* +  \sum_{ j  \;\text{downstream of}\;  i } \delta R_j^*,
\label{eq:flux_convergence}
\end{equation}
where $j$ runs over all resources that appear as products in reactions downstream of $R_i$. Since the resource concentrations are set by thermodynamics (to leading order), the fluxes in the community also converge to this thermodynamically determined value. Thus we observe functional convergence in the community.

\section{Incorporating maintenance energy requirements}
We can incorporate the maintenance energy requirements of a cell by replacing $\delta$ in Eq.~\eqref{eq:dNdt} such that 
\begin{equation}
\frac{d N_{\alpha}}{d t}= N_{\alpha} \left[ \sum_{i,j \in \mathcal{X} } E^{(\alpha)}_{i j} \mathcal{F}_{ij} \mathcal{E}^{\mathrm{ATP}}_{ij} Y -\delta -m_{\alpha} \right],
\label{eq:dNdt}
\end{equation}
where $m_{\alpha}$ is the energy flux required for cell maintenance. If the variation in maintenance costs between species is small compared to the average maintenance cost, $m$,  the analysis of the model without maintenance costs can be repeated in a straightforward manner. 

The equations describing resource concentrations all apply with a new definition of $\mu_{ij}$, given by
\begin{equation}
\mu_{ij}=  \frac{\delta + m }{k_{cat ,ij} \mathcal{E}^{\mathrm{ATP}}_{ij} Y }  \left[\mathbf{E^{-1}}.  \overrightarrow{1}\right]_{ij},
\end{equation}
where terms capturing the species-dependent variation in maintenance energy are neglected as small corrections to $\mu_{ij}$. Once again, we find that resource concentrations are determined by thermodynamics, to leading order in $(\delta + m)/g_{max}$. From Eq.~\eqref{eq:max_dissip_mu}, we see that the maximum dissipation principle determines the network selected by ecological competition.

Again, by balancing fluxes we can derive the reaction flux into a resource $R_i$, $I_i$, as
\begin{equation}
I_i^*= \delta  R_i^* +  \sum_{ j  \;\text{downstream of}\;  i } \delta R_j^*.
\end{equation}
Note that this flux is still proportional to $\delta$ because resource loss in this model only occurs via dilution. This resource dilution is essential to ensure that the community is at a non-equilibrium steady-state.

\begin{figure}
\includegraphics[width=.5\textwidth]{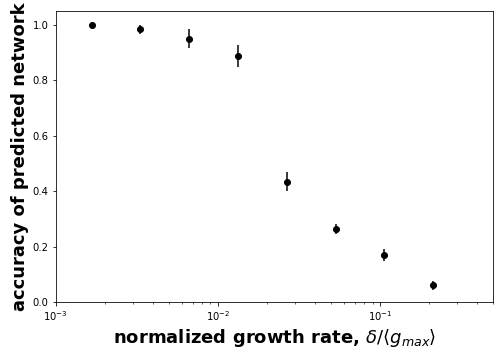}
\caption{\textbf{Maximum dissipation predicts metabolic network accurately for larger variation in enzymatic parameters.} The accuracy of the network predicted by maximum dissipation was quantified as the fraction of reactions in community metabolic network correctly predicted. The enzymatic parameters were picked from lognormal distributions with $\mu=1$ and $\sigma=0.15$. Results were average over ten species pools with enzyme budget and allocation chosen similarly as in Fig.2 in the main text. The environment was the same as in Fig.2 in the main text. }
\label{fig:high_EnzVar}
\end{figure}

\section{Generalising to arbitrary reaction stoichiometry}
In this section we generalize the model to the scenario where the stoichiometry of the substrate to product reaction is no longer $1:1$. We will show that our main results of maximum dissipation and functional convergence are robust to changes in stoichiometry. Stoichiometry only modifies the equations quantitatively.

Specifically, we consider the enzyme catalyzed reversible reaction 
\begin{equation}
E + y_s S  \rightleftharpoons   E + y_p P,
\end{equation}
where $y_s, y_p$ are the stoichiometric coefficient of the substrate and product in the reaction. The heat dissipated in the reaction, defined as  
\begin{equation}
Q_{SP}=y_s \mathcal{E}_S^0- y_p\mathcal{E}_P^0 - \mathcal{E}^{\mathrm{ATP}},
\end{equation}
accounts for reaction stoichiometry.
The reaction flux is again described by reversible Michaelis-Menten kinetics~\cite{liebermeister_modular_2010, noor_note_2013}:  
\begin{equation}
J= k_{cat} E  \frac{ \left(\frac{S}{K_S} \right)^{y_s}  }{1 +  \left(\frac{S}{K_S}\right)^{y_s} + \left(\frac{P}{K_P}\right)^{y_p}  } \left(1- \frac{P^{y_p} }{S^{y_s} } \gamma \right),
\label{eq:J_stoich}
\end{equation}
where $\gamma= e^{- Q_{SP} /RT}$ and the flux per unit enzyme being $F=J/E$.

We now consider a full connected network of resources $R_0, R_1,...R_M$, ordered in terms of their standard-state energies as before. The stoichiometric coefficients of the reaction in the network cannot be entirely arbitrary. For e.g., a mole of glucose (with 6 carbons), can produce no more than 3 moles of acetate (with 2 carbons). Motivated by this, for our reaction network, we enforce the constraint that the number of moles of resource $R_i$ produced from a single mole of $R_0$ should be independent of the reaction path. As a concrete example, the two reaction paths from $R_0$ to $R_2$ will have the reactions
\begin{align*}
 y_0 R_0  &\xrightleftharpoons{E_{01} }  y_1 R_1,  \xrightleftharpoons{E_{12}}    y_2 R_2,\\
y_0 R_0   &\rightleftharpoons{E_{02} }   y_2 R_2.
\end{align*}
Note that the stoichiometric coefficient of $R_2$ in both reactions is the same, $y_2$, due to this stoichiometric constraint. Without loss in generality, we assume that the stoichiometric coefficient of $R_0$, $y_0$ is equal to 1. This allows us to interpret the stoichiometric coefficient of each resource in the network as the number of moles produced per mole of the supplied resource $R_0$.

The equations for species and resource dynamics extend in a straightfoward manner as:

\begin{equation}
\frac{d N_{\alpha}}{d t}= N_{\alpha} \left[ \sum_{i,j \in \mathcal{X} } E^{(\alpha)}_{i j} \mathcal{F}_{ij} \mathcal{E}^{\mathrm{ATP}}_{ij} Y  -\delta \right],
\label{eq:dNdt_stoich}
\end{equation}
and 
\begin{equation}
\frac{d R_i}{d t}= h(R_i) - \sum_{\alpha, j} y_i N_{\alpha} E^{(\alpha)}_{i j} \mathcal{F}_{ij}+ \sum_{\alpha, k} y_i N_{\alpha} E^{(\alpha)}_{ki} \mathcal{F}_{ki} - \delta R_i.
\label{eq:dRdt_stoich}
\end{equation}

We can again solve for the steady-state by using the inverse of the enzyme allocation matrix of the survivors as above. The analog of Eq.~\eqref{eq:matrix_Inv} will now be 
\begin{equation}
\left(1-\gamma_{ij} \frac{ \left( R^*_j \right)^{y_j}}{ \left( R^*_i \right)^{y_i} }\right)  =\frac{\delta}{k_{cat ,ij} \mathcal{E}^{\mathrm{ATP}}_{ij} Y }   \frac{1+ \left( \frac{R^*_i}{K_{S_{ij}}}\right)^{y_i} + \left(\frac{R^*_j}{K_{P_{ij}}  } \right)^{y_j} }{ \left( \frac{R^*_i}{K_{S_{ij}}} \right)^{y_i}  } \left[\mathbf{E^{-1}}  \overrightarrow{1}\right]_{ij} ,
\label{eq:matrix_Inv_stoich}
\end{equation}

Like above, we can simplify this in the high-flux limit (Eq. \eqref{eq:highflux}), where reaction fluxes are primarily slowed down by thermodynamic inhibition. For each active reaction in the final community, we get the analog of Eq.~\eqref{eq:RjRi_ss_matrix} : 

\begin{equation}
 \frac{ \left( R^*_j \right)^{y_j}}{ \left( R^*_i \right)^{y_i} }= \frac{1}{\gamma_{ij}}  \left( 1-   \frac{\delta}{k_{cat ,ij} \mathcal{E}^{\mathrm{ATP}}_{ij} Y }  \left[\mathbf{E^{-1}}  \overrightarrow{1}\right]_{ij} \right).
\label{eq:RjRi_ss_matrix_stoich}
\end{equation}
Thus $ \frac{ \left( R^*_j \right)^{y_j}}{ \left( R^*_i \right)^{y_i} }$ is set by thermodynamics to leading order in $\delta$. 

To simplify the notation, we will define $\mu_{ij}$ once again as
\begin{equation}
\mu_{ij}=  \frac{\delta}{k_{cat ,ij} \mathcal{E}^{\mathrm{ATP}}_{ij} Y }  \left[\mathbf{E^{-1}}.  \overrightarrow{1}\right]_{ij} 
\end{equation}

We can use the same procedure as above to express the concentration ratio of two resources for an arbitrary tree-like metabolic network in terms of the heat dissipated on the path between the two resources. 
\begin{equation}
 \frac{ \left( R^*_j \right)^{y_j}}{ \left( R^*_i \right)^{y_i} }=
\prod_{u,v \in \mathcal{P}_{ij}} \frac{1}{\gamma_{uv}}  \left( 1-   \mu_{uv} \right),
\end{equation}
where $\mathcal{P}_{ij}$ is set of reactions on the path from $R_i$ to $R_j$. To leading order in $\delta$ we get : 
\begin{equation}
   \frac{ \left( R^*_j \right)^{y_j}}{ \left( R^*_i \right)^{y_i} }= e^{Q_{i \Rightarrow j}/RT}  \;  + \mathcal{O}\left(\frac{\delta}{g_{\mathrm{max}}}\right).
    \label{eq:RiRj_at_SS_stoich}
\end{equation}
This equation shows that the ratio of steady-state resource concentrations (raised to powers based on stoichiometric coefficients) on an arbitrary tree network is once again determined by the heat dissipated on the path between the two resources.

We now study the question of which steady-state network gets selected as a result of ecological competition between species. Specifically, we consider the outcome of competition between species using different paths to the same resource, $R_k$. Like before, for a given steady state community, we consider the invasion of species $\beta_{ik}$ specializing on a reaction $R_i \to R_k$ that is inactive in the current community (Fig.~\ref{fig:SI_cartoon_invasion}). For the invader to succeed, we need it to have a positive growth rate in the environment it is invading, i.e.,
\begin{equation}
k_{cat ,ik}E^{(\beta_{ik})}_{ik} Y\mathcal{E}^{\mathrm{ATP}}_{ik} \left(1- \gamma_{ik} \frac{ \left(R_k^*\right)^{y_k} }{\left(R_i^*\right)^{y_i}}\right)  - \delta>0,
\end{equation}
where $R_k^*, R_i^*$ are the steady-state resource concentrations in the community prior to invasion. Using  (Eq.\eqref{eq:RiRj_at_SS_stoich}), we can simplify this invasion condition as

\begin{equation}
\frac{\prod_{u,v \in \mathcal{P}_{0k}} \frac{1}{\gamma_{uv}}  \left( 1-   \mu_{uv} \right) }{\prod_{u\prime,v\prime \in \mathcal{P}_{0i}} \frac{1}{\gamma_{u\prime v\prime}}  \left( 1-   \mu_{u\prime v\prime} \right)  } < \frac{1}{\gamma_{ik}}  \left( 1-   \mu_{ik} \right),
\end{equation}
where the paths $ \mathcal{P}_{0k}, \mathcal{P}_{0i} $ refer to the reaction paths in the community before invasion. For small enough $\delta$ such that $\mu_{ij}= \frac{\delta }{k_{cat ,ij} \mathcal{E}^{\mathrm{ATP}}_{ij} Y E^(\beta)_{ij}  } \ll 1$ for any reaction, the above expression simplifies to 

\begin{equation}
Q_{0\rightarrow k}< Q_{0\rightarrow i} +Q_{i k},
\end{equation}
we have used the definition of $\gamma$.  $Q_{0\rightarrow k}$ and $Q_{0\rightarrow i}$ refer to the total heat dissipated over the reactions on the paths $R_0 \rightarrow R_k$ and $R_0 \rightarrow R_i$ in the steady state community before invasion, and $Q_{i k}$ is the heat dissipated in the reaction utilized by the invader. Hence the RHS equals the heat dissipated on the new path from $R_0$ to $R_k$ created by the invader. Thus the invader is able to successfully invade only if the new path to $R_k$ dissipates more heat than the path to $R_k$ in the resident community.  So once again we discover that the principle of maximum dissipation determines the reaction network selected by ecological competition of many species.

We now address the question of functional convergence. Flux conservation in the reaction network gives us
\begin{equation}
\sum_{i=0}^M y_i R^*_i \delta =  h_0 \delta,
\end{equation}
where $h_0$ is the concentration of the resource supply and $y_0=1$. Using Eq.~\eqref{eq:RiRj_at_SS_stoich}, we can write this in terms of only $R^*_0$:
\begin{equation}
\sum_{i=0}^M y_i \left(\frac{R^*_0}{\gamma_{0i}}\right)^{1/y_i} - h_0 =0.
\end{equation}

Unlike the $1:1$ stoichiometry scenario, here we get a nonlinear equation for $R_0$. A nonlinear equation could have multiple solutions for $R^*_0$, which would correspond to alternative steady-states of community function. However, because the stoichiometric coefficients $y_i$ are all positive we are guaranteed to have only a single positive solution for $R_0^*$ by Descartes rule of signs (extended to rational exponents)~\cite{curtiss_recent_1918}. We can also understand this intuitively by noticing that the function is monotonically increasing for positive $R^*_0$. Thus we have a single positive solution for $R^*_0$, which we can obtain numerically.

Using the solution for $R^*_0$,Eq.~\eqref{eq:RiRj_at_SS_stoich}, and flux conservation at each resource node, we can calculate the fluxes in each reaction as given by Eq.~\eqref{eq:flux_convergence}. Thus we have functional convergence for arbitrary reaction stoichiometries.

\section{A model without resource dilution}

We now consider a model without resource dilution. Without any flow of resource out of the system, the system cannot reach a non-trivial steady-state---the constant inflow of resources will eventually clog up the system. Here, we show that an effective outflow of resources can appear from anabolism accompanied by species death. Again, the principle of maximum dissipation determines the community metabolic network.

In this model, resources are used to maintain biomass through anabolic processes by species at resource-specific rates $b_i$. A species $\alpha$ dies or sinks with rate $m_\alpha$ . The accompanying loss of species biomass leads creates a flux of resource out of the system, which allows the system to attain a non-equilibrium steady-state.

The equation describing dynamics of species $\alpha$  with abundance  $N_{\alpha}$ is
\begin{equation}
\frac{d N_{\alpha}}{d t}= N_{\alpha} \left[ \sum_{i,j} E^{(\alpha)}_{i j} \mathcal{F}_{ij}  Y    -m_\alpha \right],
\end{equation}
where we $\mathcal{F}$ denotes the per-capita flux as defined previously.

The dynamics of resource $i$  with concentration $R_i$ is given by
\begin{equation}
\frac{d R_i}{d t}= h(R_{i}) + \sum_{\alpha, k} N_{\alpha} E^{(\alpha)}_{ ki} \mathcal{F}_{ki}   - 
                \sum_{\alpha, j} N_{\alpha} E^{(\alpha)}_{i j} \mathcal{F}_{ij}  - \sum_{\alpha} N_{\alpha} b_i R_i=0,
\end{equation}
where the last term accounts for resource-assimilation for biomass by maintenance at resource-dependent rates $b_i$ and the first two terms account for reactions where $R_i$ acts as the product and substrate respectively. We assume that $b_i$ is very small and the resource needs for the slow growth and maintenance will be small.

Again, we assume only a single resource, $R_0$ is supplied with $h(R_{0})= h_0/\tau $, with $\tau$ setting the time scale of supply to allow $h_0$ to retain the dimensions of resource concentration as before.

For convenience and clarity, we study a system of specialist species . At steady-state, species growth needs to balance loss due to maintenance and/or sinking. Therefore, we have 
\begin{equation}
\frac{R_i^*}{R_j*}=\frac{1}{\gamma_{ij}}\left( 1-\frac{m_{\beta} }{k_{cat ,ij} \mathcal{E}^{\mathrm{ATP}}_{ij} Y E^{(\beta)}_{ij} } \right),
\label{eq:resource_ratio_model2}
\end{equation}
for every active reaction in the community. $\beta$ is the specialist species catalyzing reaction between $R_i$ and $R_j$.

For slow-growing communities, $m_{\beta}$ is small. Therefore,  the ratio  $\frac{R_i^*}{R_j*}$  depends only on the heat dissipated in the corresponding reaction, $\gamma_{ij} $, to leading order.

Repeating the analysis applied to the previous model, we find again that the ratio $\frac{R_i^*}{R_0*}$ depends only the heat dissipated along the realized reaction path from $R_0$ to $R_i$. Similarly, we also find that the principle of maximal dissipation determines the set of realized reactions in the network.

To arrive at the steady state concentration of resource, we use flux conservation, which reads as
\begin{equation}
\frac{h_0}{\tau}- \sum_{\alpha, i} N^*_{\alpha} b_i R^*_i=0,
\end{equation}
or rather,
\begin{equation}
R_0^*= \frac{h_0}{\tau b_0  N^*_{\mathrm{tot}}}     - \sum_{i=1}^{M} \frac{b_i}{b_0} R_i^*,
\end{equation}
where $N^*_{\mathrm{tot}}=\sum_{\alpha } N^*_{\alpha}$ is the total biomass in the community.

Thus, we arrive at the solution 
\begin{equation}
R_0^*=   \frac{ \frac{h_0}{\tau b_0 N^*_{\mathrm{tot}} }  }{1+ \sum_{i=1}^M \frac{b_i}{b_0}  e^{Q_{0\Rightarrow i}/RT} },
\end{equation}
and
\begin{equation}
R_i^*=  R_0^* e^{Q_{0\Rightarrow i}/RT},
\end{equation}
in the limit of slow growth, where $e^{-\Delta G_{0\Rightarrow i}}$ is the heat dissipated along the path from $R_0$ to $R_i$. Note that these solutions are not closed form; the solution depends on the total species biomass, $N^*_{\mathrm{tot}}=\sum_{\alpha } N^*_{\alpha}$. However, as long as the total biomass of communities assembled from different species pools in replicate environments are similar, we expect the resource concentrations in the environments to be the same. Also note that if $b_i$ is large, then substrate depletion can be a major form of competition between species, and our assumption that reactions are slowed primarily by thermodynamic inhibition does not hold.

The net flux into a resource $R_i$ is then given by 
\begin{equation}
I_i^*= N^*_{\mathrm{tot}}  b_i R_i^* +  \sum_{ j  \;\text{downstream of}\;  i } N^*_{\mathrm{tot}} b_j R_j^*
\end{equation}

Thus we have approximate functional convergence as well, with a new expression for the reaction fluxes. The functional convergence is now sensitive to variation in the total species abundance, across replicates.

\begin{figure}
\includegraphics[width=\linewidth]{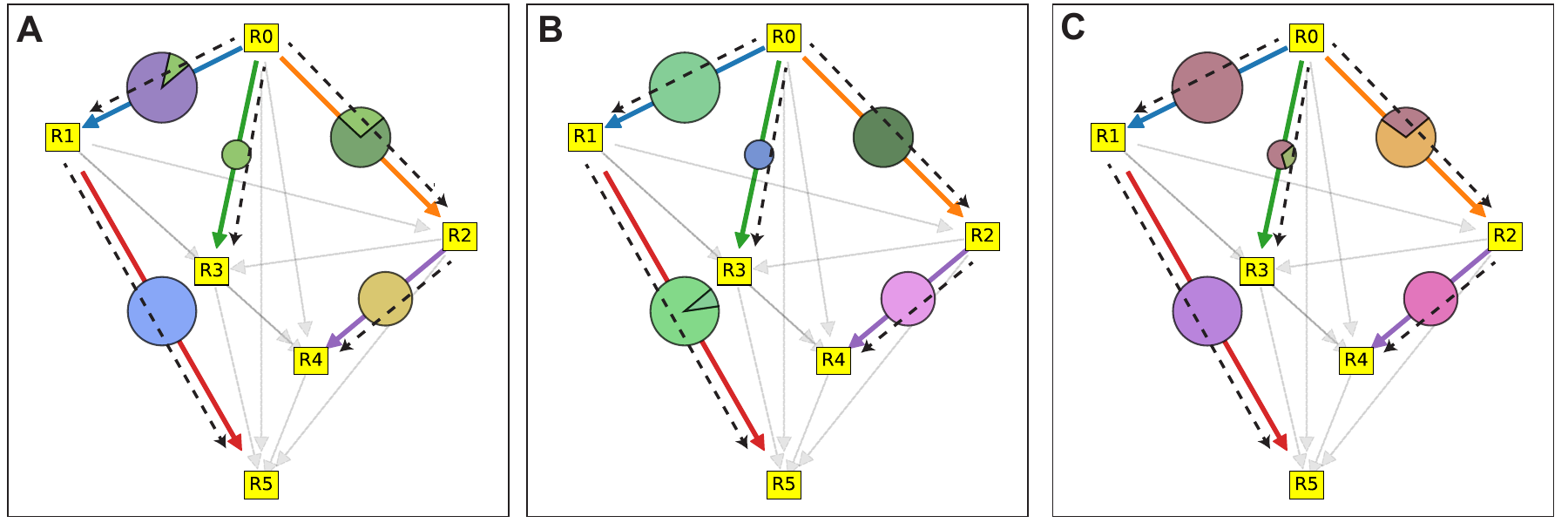}
\caption{\textbf{Maximum dissipation principle and functional convergence in model without resource dilution.}  The panels show the metabolic network of three communities assembled from three different species pools in identical environments. The maximum dissipation principle predictions, shown by dashed lines, predicts the reactions accurately. The fluxes through the reactions, represented by the size of the circles on the corresponding arrows, are similar across the three communities. The pie-charts depict how the flux is shared between species in the different communities. This differs between communities. Parameters for the figure were:  $b_i$ was chosen from as $10^-3$ times a lognormal random variable with parameters $\mu=0, \sigma=0.01$. The maintenance cost was $m=0.01$, resource supply was described by $h_0=80$ and $\tau=10$. Resource energies are the same as in Fig.2 in the main text. The energy assimilated and enzymatic parameters are chosen from the same distributions as in Fig.2 in the main text. } 
\label{fig:anabolism_noRdilution} 
\end{figure}

\bibliography{bib_thermo}
\bibliographystyle{naturemag}